\documentclass[conference, 10pt]{IEEEtran}

\usepackage{bm}
\usepackage[usenames,dvipsnames]{xcolor}
\usepackage{srcltx}
\usepackage{psfrag}
\usepackage{epsfig}
\usepackage{graphicx}
\usepackage{epstopdf}
\usepackage{color}
\usepackage[tbtags,sumlimits,nointlimits,reqno]{amsmath}
\usepackage{amssymb}
\usepackage{color}
\usepackage{float}
\usepackage{subfigure}
\usepackage{gensymb}

\usepackage{tikz}					  			
\usetikzlibrary{shapes,arrows}	          
\usetikzlibrary{shadows,arrows.meta,positioning,backgrounds,fit}

\usepackage{float}
\usepackage{mathrsfs}		         
\usepackage{mathtools}				
\usepackage{hyperref}

\begin{document}

\pagenumbering{arabic}

\title{Bessel Beams: Unified and Extended Perspective}


\author{\IEEEauthorblockN{Oscar C\'{e}spedes Vicente}
	\IEEEauthorblockA{
		Polytechnique Montr\'{e}al \\
		Montr\'{e}al, Qu\'{e}bec, H3T 1J4, Canada}
	\and
	\IEEEauthorblockN{Christophe Caloz}
	\IEEEauthorblockA{
		KU Leuven\\
		 Kasteelpark Arenberg 10 - box 2444 3001 Leuven }}




\maketitle


%
\IEEEpeerreviewmaketitle

\begin{abstract}
	We present a unified and extended perspective of Bessel beams, irrespective to their orbital angular momentum (OAM) -- zero, integer or noninteger -- and mode -- scalar or vectorial, and LSE/LSM or TE/TM in the latter case. The unification is based on the integral superposition of constituent waves along the angular-spectrum cone of the beam, and allows to describe, compute, relate, and implement all the Bessel beams, and even other types of beams, in a universal fashion. The paper first establishes the integral superposition theory. Then, it demonstrates the previously unreported existence of noninteger-OAM TE/TM Bessel beams, compares the LSE/LSM and TE/TM modes, and establishes useful mathematical relations between them. It also provides an original description of the position of the noninteger-OAM singularity in terms of the initial phase of the constituent waves. Finally, it introduces a general technique to generate Bessel beams by an adequate superposition of properly tuned sources. This global perspective and theoretical extension may open up new avenues in applications such as spectroscopy, microscopy, and optical/quantum force manipulations.
\end{abstract}

\maketitle

\vspace{5mm}
\section{Introduction}
\begin{figure}
	\tikzstyle{set} = [rectangle, rounded corners, minimum width=3cm, minimum height=0.7cm, text centered, text width=3.5cm, draw=black, line width=0.5mm, fill=white]
	\tikzstyle{subset} = [rectangle, rounded corners, minimum width=3cm, minimum height=0.7cm, text centered, text width=3.5cm, draw=black, fill=white]
	\tikzstyle{subset1} = [rectangle, rounded corners, minimum width=3cm, minimum height=0.7cm, text centered, text width=4.2cm, draw=black, fill=white]
	\tikzstyle{subset2} = [rectangle, rounded corners, minimum width=3cm, minimum height=0.7cm, text centered, text width=5.8cm, draw=black, fill=white]
	\tikzstyle{arrow} = [thick,->,>=stealth]
	
	\centering
	\begin{tikzpicture}[node distance=1.4cm]
	\node (in1) [set] {Localized Waves};
	\node (in2) [set, below of=in1] {Scalar Bessel Beam};
	\node (in3) [subset, below of=in2] {No-OAM $\qquad\qquad\quad J_0(\alpha\rho)$};
	\node (in4) [subset1, below of=in3] {Integer-OAM \mbox{$J_n(\alpha\rho)e^{in\phi},\quad n\in\mathbb{Z}$}};
	\node (in5) [subset2, below of=in4, yshift=-0.3cm] {Noninteger-OAM \mbox{$\displaystyle \sum_{m=-\infty}^{+\infty}c_m(\nu) J_m(\alpha\rho)\text{e}^{im\phi},\quad \nu\in\mathbb{R}/\mathbb{Z}$}};
	\node (in6) [set, below of=in5, yshift=-0.3cm] {Vectorial Bessel Beam};
	\node (in7) [subset, below of=in6, xshift=-2.1cm] {LSE and LSM Modes};
	\node (in8) [subset, below of=in6, xshift=+2.1cm] {TE and TM Modes};
	
	\draw [arrow] (in1) -- (in2);
	\draw [arrow] (in2) -- (in3);
	\draw [arrow] (in3) -- (in4);
	\draw [arrow] (in4) -- (in5);
	\draw [arrow] (in5) -- (in6);
	\draw [arrow] (in6) -- (in7);
	\draw [arrow] (in6) -- (in8);
	\draw [densely dotted, arrow] (in7) -- (in8);
	\draw [densely dotted, arrow] (in8) -- (in7);
	
	\end{tikzpicture}
	\caption{Classification of electromagnetic Bessel beams.}
	\label{fig:EvolutionBesselWaves}
\end{figure}
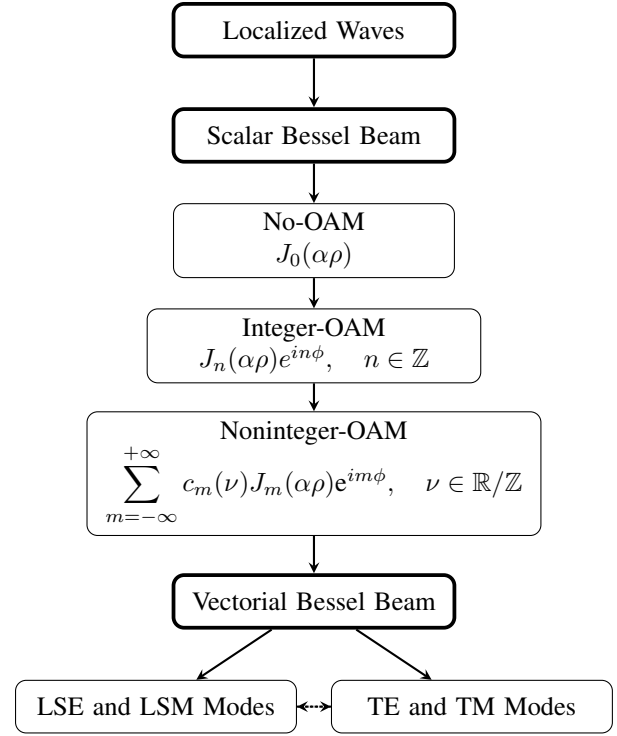

Electromagnetic Bessel beams represent a fundamental form of structured light. They are \emph{localized waves}~\cite{hernandezlocalizedone,hernandeznontwo} with transverse Bessel function profiles that carry \emph{orbital angular momentum (OAM)} along their propagation axis. \emph{Localized waves} were first reported as soliton-like waves by Bateman in 1915~\cite{bateraenelectrical}, next derived for the Bessel case as the TE/TM solutions to the cylindrical wave equation by Stratton in 1941~\cite{strattonelectromagnetic}, and then generalized as solutions to a class of equations admitting ``waves without distortion'' as solution by Courant and Hilbert in 1966~\cite{courantmethods}. They are waves that propagate without spatial dispersion (or diffraction) and without temporal (or chromatic) dispersion. Their energy is thus uniformly confined and invariant perpendicular to and along, respectively, their direction of propagation. The \emph{OAM} is a beam property whose macroscopic manifestation is an isophase surface that has the form of a vortex along the axis of the beam. It may be integer or noninteger. In the former case, the wave has the transverse phase dependence $\text{e}^{in\phi}$ ($n\in\mathbb{Z}$), corresponding to an OAM of $n\hbar$ per photon\footnote{OAM waves, or optical vortices, behave to some extent as charged particles: They may repel and attract each other, and mutually annihilate upon colliding~\cite{soskin1997topological}. For this reason, $n$ is also called the topological charge.}, while in the latter case the wave is made of a superposition of integer-OAM waves that combine so as to produce a noninteger-OAM per photon~\cite{khonina2001analysis}. This property association of localization and OAM confers to Bessel beams specific capabilities for manipulating light that may be exploited in diverse applications, such as nanoparticle guiding~\cite{arlt2001optical}, orbiting and spinning~\cite{mitrireverse}, trapping~\cite{zhan2004trapping, cizmaroptical} and tracting~\cite{mitrisingle, mitriopticalPol, mitriopticalTra}, spectroscopy~\cite{dorn2003sharper}, microscopy~\cite{planchon2011rapid} and quantum key distribution~\cite{nape2018self}. Figure~\ref{fig:EvolutionBesselWaves} shows a general classification of Bessel beams that pertain to the sequel of the paper.

\emph{Bessel beams} are the simplest form of light OAM after the Laguerre-Gauss beams and the most studied localized waves. They are monochromatic beams with a transverse amplitude pattern that follows Bessel functions of the first kind, $J_n (\alpha{\rho})$, multiplied by the phase function $\text{e}^{in\phi}$ ($n\in\mathbb{Z}$), or combinations of such waves in the noninteger OAM case. Their simplest forms are the \emph{scalar} Bessel beams, also existing in acoustics and restricted to the paraxial approximation in optics. Such beams were first experimentally showed by Durmin in 1987~\cite{durmin1987exact} for $n=0$ (no-OAM), while their integer-OAM carrying versions were reported shortly thereafter~\cite{shimoda1991exact}. Noninteger-OAM scalar Bessel beams were reported only 15 years later~\cite{gutierrez2007nondiffracting}. In the case of electromagnetic waves, such as light, Bessel beams are \emph{vectorial}, and may be either LSE/LSM or TE/TM, as mentioned in Fig.~\ref{fig:EvolutionBesselWaves}. Vectorial Bessel beams, the only exact forms of electromagnetic Bessel beams, were introduced in LSE/LSM\footnote{The terminology LSE/LSM (longitudinal section electric/magnetic) is borrowed from the theory of waveguides~\cite{balanis2016antenna}. Such modes are often referred to a linearly polarized beams in free-space optics~\cite{wang2014vector}, but the terminology LS is more rigorous since such fields also include a longitudinal component.} integer-OAM form in~\cite{mishra1991vector}~(51 years after their TE/TM form introduction by Stratton~\cite{strattonelectromagnetic}), and generalized to the noninteger case in~\cite{mitri2011}. Finally, Bessel beams of various complexities have been generated by an axicon illuminated with a Laguerre-Gaussian beam~\cite{arlt2000generation}, by a single helical axicon (azimuthal phase plate)~\cite{wei2015generation}, by a spatial light modulator~\cite{chattrapiban2003generation,gutierrez2007nondiffracting}, by a leaky-wave antenna~\cite{ettorregeneration}, and by a metasurface~\cite{pfeiffer2014controlling}.

This paper fills up some fundamental gaps existing in the literature on Bessel beams. Specifically, it reports 1)~a unified representation of Bessel beams, 2)~a demonstration of the existence of noninteger TE/TM OAMs Bessel beams and their detailed characterization, and 3)~a generic and efficient approach for the practical implementation of Bessel beams.

\noindent\textbf{\textit{ 1.~Unified Representation of Bessel Beams}} -- Bessel beams have been described either in the spatial direct ($\mathbf{r}$) domain~\cite{strattonelectromagnetic, mishra1991vector, shimoda1991exact, mitri2011} or in the spatial inverse ($\mathbf{k}$) -- or spectral -- domain~\cite{durmin1987exact,vasararealization,gutierrez2007nondiffracting}. While the spatial approach immediately describes the nature of the beam, it is restricted to the beam in question. In contrast, the spectral approach, although less explicit, allows generalizations and provides insight into simple generation techniques. The spectral approach is therefore more powerful. However, it has only been applied to the scalar case in the context of Bessel beams. This paper presents a general electromagnetic vectorial spectral formulation that applies to all Bessel beams (scalar and vectorial, with integer and noninteger OAM, LSE/LSM and TE/TM, and all related combinations), and even to other types of conical (e.g., Mathieu, Weber) and nonconical beams (e.g., Gauss-Laguerre and Hypergeometric Gaussian beams), hence providing novel perspectives and possibilities.

\noindent\textbf{\textit{ 2.~Existence and Characterisation of TE/TM with non-integer OAMs}} -- Whereas vectorial TE/TM Bessel beams~\cite{strattonelectromagnetic, ettorregeneration, shimoda1991exact, feshbach2019methods} and scalar Bessel beams with arbitrary OAM~\cite{gutierrez2007nondiffracting,yangnondiffracting} have been separately described in the literature, we report here for the first time the existence of TE/TM Bessel beams with non-integer OAMs, and present a detailed description of these new modes, showing superior focusing capabilities~\cite{quabis2000focusing} than their LSE/LSM counterparts in addition to richer optical force opportunities.

\noindent\textbf{\textit{ 3.~Generic and Efficient Practical Implementation}} -- Both scalar and vectorial Bessel beams have been experimentally generated using diverse technologies, all suffering from some drawbacks. The technologies for scalar beams include the helical axicon~\cite{wei2015generation}, which has the disadvantage of being a bulky lens, and spatial light modulators~\cite{chattrapiban2003generation,gutierrez2007nondiffracting}, which have restricted polarization flexibility due to their inherent phase-only and magnitude-only restriction. The technologies for vectorial beams include a leaky-wave antennas~\cite{ettorregeneration}, which can produce specific low and integer-order OAM TE/TM beams, and metasurfaces~\cite{pfeiffer2014controlling}, which are the most attractive technology but which have been conceived in the direct domain and have hence not yet benefited from the generality of the vectorial spectral-domain formulation~\footnote{This formulation corresponds to an extended ($\nu\in\mathbb{R}$) scalar-spectrum formulation in~\cite{hernandezlocalizedone,hernandeznontwo} and is complementary to the method of carrying out a Fourier transform on a scalar Bessel beam with a Gaussian envelope~\cite{chafiqoptical} (see~\hyperref[app:BBs_vs_BBswithGBenv]{Supp. Mat. H})} presented in this paper. Based on this approach, we present here an efficient and generic approach for the practical implementation of any Bessel beam (and other conical or nonconical beams).

\section{\label{sec:level2}Scalar Solution}

\begin{figure}[h!]
	\centering
	\includegraphics[scale=1]{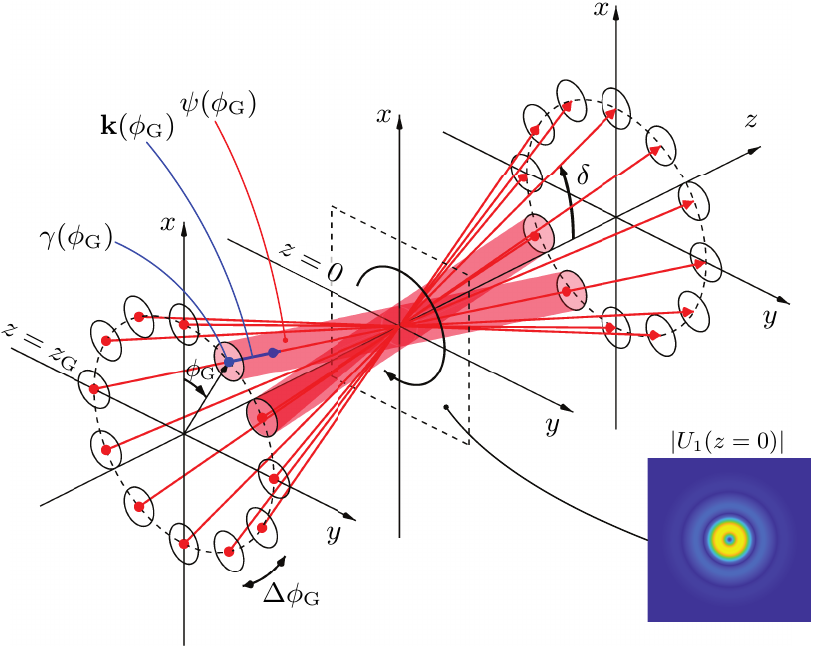}
	\caption{General description of a Bessel beam [Eq.~\eqref{eq:ScalarIntegralFormPsi}] as a superposition of waves [Eq.~\eqref{eq:Psi_Gen_wave}] with different phases [Eq.~\eqref{eq:Lin_Phase}] propagating along a cone of opening angle $\delta=\tan^{-1}(\alpha/\beta)$ [Eq.~\eqref{eq:Conical_k}] towards its apex. The field panel shows, as an example, $|U_1(x,y,0,0)|=|J_1(\alpha\rho)|$.}
	\label{fig:SpectrumDoubleConePart1}
\end{figure}

A Bessel beam may be generally described in a linear system by the scalar function of space ($x$, $y$, $z$) and time ($t$)
\begin{equation}
	U_{\nu}(x,y,z,t)=\int_{0}^{2\pi}\psi_\nu(\phi_\text{G})d\phi_\text{G},
	\label{eq:ScalarIntegralFormPsi}
\end{equation}
where $\psi_\nu(\phi_\text{G})$ represents a continuous set of waves that propagate at the angle $\phi_\text{G}$ along a circular cone towards its apex, so as to form a transverse interference pattern in the form of a Bessel function, as illustrated in Fig.~\ref{fig:SpectrumDoubleConePart1}, and where $\nu$ is equal to the OAM of the beam when it is integer or half-integer (see~\hyperref[app:OAM_Bessel_Global_Order]{Supp. Mat. A}). These waves have the mathematical form
\begin{subequations}
	\begin{equation}\label{eq:Psi_Gen_wave}
		\mathbf{\psi_\nu(\phi_{\text{G}})}=\text{e}^{i(\mathbf{k}(\phi_{\text{G}})\cdot\mathbf{r}+\gamma_\nu(\phi_{\text{G}})-\omega t)}w(\xi),
	\end{equation}
	with the oblique wave vector
	\begin{equation}
		\mathbf{k}(\phi_{\text{G}})= -\alpha\left(\cos(\phi_{\text{G}})\hat{\mathbf{x}}+\sin(\phi_{\text{G}})\hat{\mathbf{y}}\right)+\beta\hat{\mathbf{z}},
		\label{eq:Conical_k}
	\end{equation}
	where
	\begin{equation}
		\alpha=k_0\sin(\delta),\quad\beta=k_0\cos(\delta),
		\label{eq:del_alphabeta}
	\end{equation}
	so that $\alpha^2+\beta^2=k_0^2=\left(\omega/c\right)^2$, and $\delta$ is called the axicon angle, and with the linear phase	
	\begin{equation}
		\gamma_\nu(\phi_\text{G})=2\pi\nu\ \text{frac}\left(\frac{\phi_\text{G}}{2\pi}+\left(1-\frac{\phi_{\text{G},0}}{2\pi}\right)\right),\quad \nu \in \mathbb{R}.
		\label{eq:Lin_Phase}
	\end{equation}
	\label{eq:Psi_Gen}
\end{subequations}
In these relations, $w(\xi)$ is the transverse apodization of each wave with respect to its propagation direction, $\mathbf{k}(\phi_{\text{G}})$, with $\xi$ being the radial variable of the corresponding local coordinate system, $\text{frac}(\cdot)$ is the fractional part function~\cite{WeissteinFrac}, and $\phi_{\text{G},0}$ is related to the position of the phase singularity for $\nu\in\mathbb{R}\backslash\mathbb{Z}$, as will be seen later. Note that Eq.~\eqref{eq:Lin_Phase} represents a sawtooth function of $\phi_{\text{G}}$ that reduces to $\gamma_\nu(\phi_\text{G})=\nu\phi_{\text{G}}$ for $\phi_{\text{G}}\in[0,2\pi[$ and $\phi_{\text{G},0}=0$. Also note that possible loss may be simply accounted for by making the $\mathbf{k}$ vector complex.

Practically, the waves $\mathbf{\psi_\nu(\phi_{\text{G}})}$ must be spatially limited (e.g. Gaussian cross-sectional apodization $w(\xi)$), but they will be initially considered as plane waves ($w(\xi)=\text{const.}$)\footnote{Note that the beam is still localized in this case due to the decreasing envelope of the Bessel interference pattern.} because, as we shall see, an analytical solution can be derived for Eq.~\eqref{eq:ScalarIntegralFormPsi} in this case\footnote{Eq.~(\ref{eq:ScalarIntegralFormPsi}) may then be alternatively expressed as the inverse Fourier transform $U_\nu(\rho,\phi,z)=\mathcal{F}^{-1}\{\tilde{U}_\nu(k_\rho,k_\phi,k_z)\}=\int_0^{\infty}\int_{0}^{2\pi}\int_0^{\infty}\tilde{U}_\nu(k_\rho,k_\phi,k_z)\text{e}^{i(\boldsymbol{k}\cdot\mathbf{r}-\omega t)}k_\rho dk_\rho dk_\phi dk_z$ with the spectrum
	$\tilde{U}_\nu(k_\rho,k_\phi,k_z)=
	\frac{\delta(k_\rho-\alpha)}{k_\rho}\text{e}^{i\nu(k_\phi+(2\pi-(\phi_{\text{G},0}\pm\pi)))}\delta(k_z-\beta) \text{ if }k_\phi<\phi_{\text{G},0}\pm\pi$ and $\tilde{U}_\nu(k_\rho,k_\phi,k_z)=
	\frac{\delta(k_\rho-\alpha)}{k_\rho}\text{e}^{i\nu(k_\phi-(\phi_{\text{G},0}\pm\pi))}\delta(k_z-\beta) \text{ if }k_\phi\geq\phi_{\text{G},0}\pm\pi$ (see~\hyperref[app:Ang_Spectrum]{Supp. Mat. B}).}. Moreover, their azimuthal separation ($\Delta\phi_{\text{G}}$) is infinitesimally small, so that they effectively merge into the azimuthal continuum corresponding to the integral~\eqref{eq:ScalarIntegralFormPsi}. Finally, these waves may be temporarily considered as a scalar, but they will later be seen to represent any of the components of fully vectorial electromagnetic waves.

\begin{figure}[h]
	\centering
	\includegraphics[scale=1]{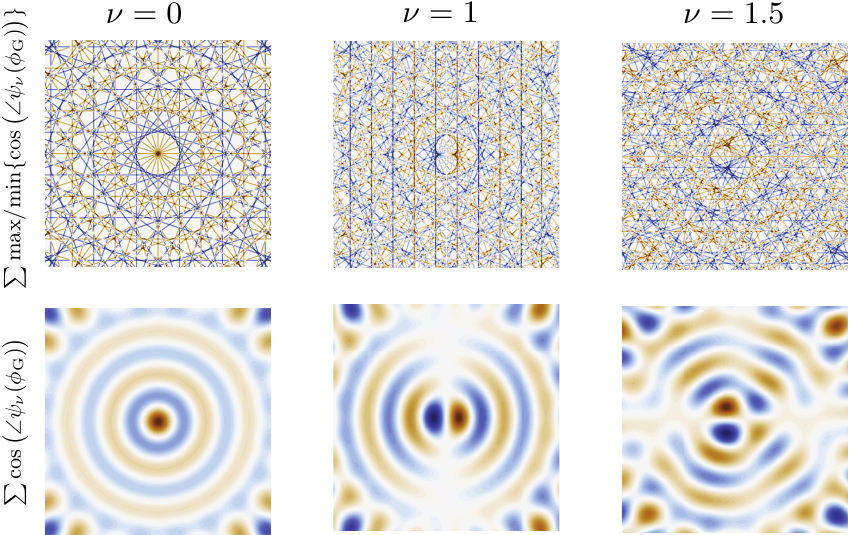}
	\caption{Cross-sectional view of the beam numerically obtained by discretizing the integral in Eq.~\eqref{eq:ScalarIntegralFormPsi} with 20 equi-spaced constituting plane waves [Eq.~\eqref{eq:Psi_Gen_wave} with $w(\xi)=1$] in the plane $z=0$ of Fig.~\ref{fig:SpectrumDoubleConePart1} for $\nu=0,1,1.5$ at $t=0$ and over the cross-sectional area of $7\lambda\times 7\lambda$ ($\lambda=2\pi c/\omega$) co-centered with the beam axis. Top Row: Maxima (red lines) and minima (blue lines). Bottom Row: Actual waves.}
	\label{fig:MinMaxAmplitude}
\end{figure}

The Bessel beam superposition in Eq.~\eqref{eq:ScalarIntegralFormPsi} is plotted in Fig.~\ref{fig:MinMaxAmplitude} using 20 equi-spaced plane waves for $\mathbf{\psi_\nu(\phi_{\text{G}})}$ in Eq.~\eqref{eq:Psi_Gen}, with the top row showing the superposition of the maxima and minima, and the bottom row plotting the superposition of the actual waves with continuous sinusoidal gradients. This figure shows that a perfectly smooth Bessel pattern is obtained around the axis of the beam with a restricted number of constituent waves\footnote{The top row of Fig.~\ref{fig:MinMaxAmplitude} shows that this radial effect is due to the radially decreasing density of the constituent waves.}, and illustrates the increasing structural complexity of the Bessel beam without OAM ($\nu=0$), with integer OAM ($\nu=1$) and with non-integer OAM ($\nu=1.5$).

Substituting Eq.~(\ref{eq:Psi_Gen}) with $w(\xi)=A^{\text{PW}}$ (const.) into Eq.~(\ref{eq:ScalarIntegralFormPsi}), simplifying the resulting integral (see~\hyperref[app:Concise_Form]{Supp. Mat. C}), and decomposing the field into its transverse-dependence and longitudinal/temporal-dependence parts as $U_{\nu}(\rho,\phi,z,t)=U_{\nu,\perp}(\rho,\phi)\text{e}^{i(\beta z - \omega t)}$, yields then
\begin{equation}
	U_{\nu,\perp}(\rho,\phi)=A^{\text{PW}}\int_{\phi_{\text{G},0}}^{\phi_{\text{G},0} +2\pi} \text{e}^{-i \alpha \rho \cos(\phi'-\phi)}\text{e}^{i\nu(\phi'-\phi_{\text{G},0})} d\phi'.
	\label{eq:TransScalInt}
\end{equation}

For $\nu=n\in\mathbb{Z}$, the integral in Eq.~(\ref{eq:TransScalInt}) has a tabulated closed-form primitive~\cite{abramowitz1988handbook}, leading to $U_{n,\perp}=2\pi i^{-n} \text{e}^{-in\phi_{\text{G},0}}A^{\text{PW}}J_n(\alpha \rho) \text{e}^{i n\phi}$, which is the conventional integer-OAM Bessel solution for circular-cylindrical problems. In contrast, for $\nu\in\mathbb{R}\backslash\mathbb{Z}$, the integral does not admit a simple primitive, and we must devise a strategy to lift this restriction so as to find the most general beam solution. This can be accomplished by the following procedure. First, we replace the generally non-periodic ($\nu\in\mathbb{R}\backslash\mathbb{Z}$) complex exponential function $\text{e}^{i\nu\phi'}$ in Eq.~(\ref{eq:TransScalInt}) by its expansion in terms of the complete and orthogonal set of \emph{periodic} ($\nu=m\in\mathbb{Z}$) complex exponential functions $\text{e}^{im\phi'}$ (see~\hyperref[app:Orthogonal_Expansion]{Supp. Mat. D})~\cite{arfken1999mathematical}, i.e.,
\begin{equation}
	\text{e}^{i\nu \phi'}=\sum_{m=-\infty}^{+\infty}\text{sinc}\big((m-\nu)\pi\big)\text{e}^{-i(m-\nu)(\phi_{\text{G},0}+\pi)}\text{e}^{i m \phi'}.
	\label{eq:Complex_Orthogonal_Expansion}
\end{equation}

Then, we substitute Eq.~(\ref{eq:Complex_Orthogonal_Expansion}) into Eq.~(\ref{eq:TransScalInt}), which leads to
\begin{equation}
	\begin{aligned}
		U_{\nu,\perp}(\rho,\phi)&=A^{\text{PW}}\text{e}^{-i\nu\phi_{\text{G},0}} \sum_{m=-\infty}^{+\infty}\text{sinc}\big((m-\nu)\pi\big)\text{e}^{-i(m-\nu)(\phi_{\text{G},0}+\pi)}\\
		&\hspace{1.2 cm}\times \int_{\phi_{\text{G},0}}^{\phi_{\text{G},0}+2\pi}\text{e}^{-i \alpha \rho \cos(\phi-\phi')}\text{e}^{i m \phi'} d\phi'.\\
	\end{aligned}
	\label{eq:InvFourierCylSpecGen}
\end{equation}

Finally, we eliminate the integral by applying the Bessel identity $\int_0^{2\pi}\text{e}^{-i \alpha \rho\cos(\phi-\phi')}\text{e}^{i m \phi'}d\phi'=2\pi i^{-m}J_m(\alpha \rho) \text{e}^{i m\phi}$~\cite{abramowitz1972handbook}, and find the general solution to Eq.~(\ref{eq:InvFourierCylSpecGen}), with the singularity phase parameter $\phi_{\text{G},0}$ appended to the corresponding expression given in~\cite{gutierrez2007nondiffracting}, in the closed form
\begin{subequations}
	\begin{equation}\label{eq:BesselAnyOAMa}
		U_{\nu,\perp}(\rho,\phi)=\sum_{m=-\infty}^{+\infty}A_m^{\text{BB}}(\nu,\phi_{\text{G},0}) J_m(\alpha\rho)\text{e}^{im\phi},
	\end{equation}
	with the complex weighting distribution
	\begin{equation}\label{eq:BesselAnyOAMb}
		A_m^{\text{BB}}(\nu,\phi_{\text{G},0})=
		2\pi i^{-m}  \text{e}^{-i\nu\phi_{\text{G},0}} A^{\text{PW}} \text{sinc}\big((m-\nu)\pi\big)\text{e}^{-i(m-\nu)\left(\phi_{\text{G},0}+\pi\right)}.
	\end{equation}
	\label{eq:BesselAnyOAM}
\end{subequations}

Equation~(\ref{eq:BesselAnyOAM}) contains the integer-OAM Bessel solution, since $\nu=n\in\mathbb{Z}$ transforms Eq.~(\ref{eq:BesselAnyOAMb}) into $A_m^{\text{BB}}=2\pi i^{-m}\text{e}^{-in\phi_{\text{G},0}}A^{\text{PW}}\delta_{mn}$, which reduces the sum in Eq.~(\ref{eq:BesselAnyOAMa}) to the single term $U_{n,\perp} = 2\pi i^{-n}\text{e}^{-in\phi_{\text{G},0}}A^{\text{PW}}J_n(\alpha\rho)e^{in\phi}$. But it also contains noninteger-OAM ($\nu\in\mathbb{R}\backslash\mathbb{Z}$) solutions, where the satisfaction of the circular periodic boundary condition is realized by a \emph{superposition} of integer-OAM Bessel waves with proper phase ($\text{e}^{im\phi}$) and weighting coefficients ($A_m^{\text{BB}}(\nu,\phi_{\text{G},0})$). In this noninteger-OAM case, the sum in Eq.~\eqref{eq:BesselAnyOAMa} must be practically truncated to an integer $m=\pm{M}$ that is large enough to provide a satisfactory approximation of the Bessel beam.

Equation~\eqref{eq:BesselAnyOAMa} reveals that the parameter $\alpha$ of the cone in Fig.~\ref{fig:SpectrumDoubleConePart1} corresponds to the spatial frequency of the Bessel pattern. Since this parameter is proportional to the axicon angle, $\delta$, according to Eq.~\eqref{eq:del_alphabeta}, we find that increasing the aperture of the cone in the integral construction of Eq.~\eqref{eq:ScalarIntegralFormPsi} compresses the Bessel ring pattern towards the axis of the beam.

\begin{figure}[ht]
	\centering
	\includegraphics[scale=0.95]{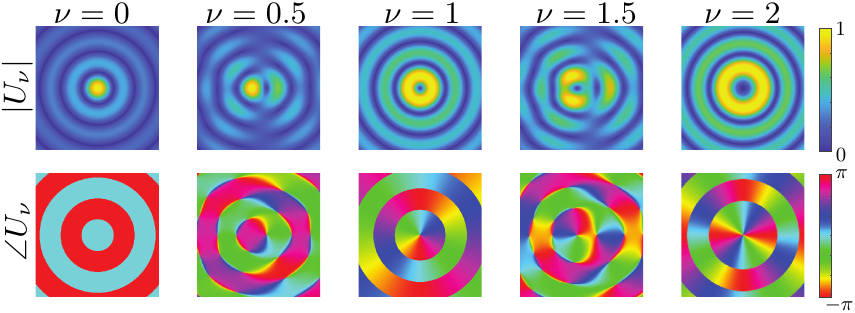}
	\caption{Cross-sectional view of the complex Bessel beam analytically computed by Eq.~(\ref{eq:BesselAnyOAM}) for different OAM orders ($\nu$), with $\phi_{\text{G},0}=0$, $\delta$ = 25$^\circ$ and $M=100$, over the same area as in Fig.~\ref{fig:MinMaxAmplitude}. Top row:~Transverse amplitude. Bottom row:~Transverse phase.}
	\label{fig:SpectrumDoubleConePart2}
\end{figure}

Figure~\ref{fig:SpectrumDoubleConePart2} plots the magnitude and phase of the Bessel beam given by Eq.~\eqref{eq:BesselAnyOAM} for different integer and noninteger OAMs~\footnote{The fractional superposition of Bessel waves produces novel beam properties, such as internal vortices~\cite{mitrireverse} and negative wave propagation~\cite{mitri2012high}, which are beyond the scope of this paper.}. The cases $\nu=0$, $1$ and $1.5$ correspond to the instantaneous field plots in Fig.~\ref{fig:MinMaxAmplitude}. The OAM-less beam $\nu=0$ has simultaneously azimuthally symmetric magnitude and phase. The integer beams $\nu=n\in\mathbb{N}$ have azimuthally symmetric magnitudes but asymmetric phase (OAM). The beams $\nu\in\mathbb{R}\backslash\mathbb{Z}$ have simultaneously asymmetric magnitude and phase. The parameter $\phi_{\text{G},0}$, which is here $0$, corresponds to a dummy initial phase of the integer-OAM  and the position of the discontinuity of the noninteger-OAM in the individual waves, with increasing $\phi_{\text{G},0}$ clockwise rotating the asymmetric magnitude of the noninteger-OAM pattern (see~\hyperref[app:NL_Discontinuity_Parameter]{Supp. Mat. E}).

\section{Vectorial Solution Construction}
The general\footnote{By the term `general', we refer here both to the arbitrariness of $\nu$ ($\nu=n\in\mathbb{N}$ or $\nu\in\mathbb{R}\backslash\mathbb{N}$) and to the alternative constructions of Eqs.~(\ref{eq:ScalarIntegralFormPsi}) (integral) and (\ref{eq:BesselAnyOAM}) (analytical), with the discontinuity parameter $\phi_{\text{G},0}$.} scalar Bessel solutions described above are restricted to acoustic waves, quantum waves, and vectorial waves under special conditions such as the paraxial condition ($\delta\ll\pi/2$) in the electromagnetic case. On the other hand, they fail to describe Bessel beams with a large axicon aperture (angle $\delta$ in Fig.~\ref{fig:SpectrumDoubleConePart1}), which are relevant to applications such as microscopy and optical force manipulations. Therefore, we extend here the previous scalar generalization to the vectorial case.

For the scalar case, we have established two alternative solutions, the integral solution of Eq.~\eqref{eq:ScalarIntegralFormPsi} and the analytical solution of Eq.~(\ref{eq:BesselAnyOAM}). In the present extension to the vectorial case, we shall restrict our treatment to the integral approach\footnote{The vectorialization of the analytical solution is no more complicated than that of the integral solution, and it is formally equivalent to the auxiliary potential vector method~\cite{ishimaru2017electromagnetic}.}, because it provides more insight into the physical nature of the beam and because it will constitute the basis for the practical implementation to be discussed later. We shall still assume a plane wave construction ($w(\xi)=A^{\text{PW}}$ (const.) in Eq.~\eqref{eq:Conical_k}), for simplicity.

As indicated in Fig.~\ref{fig:EvolutionBesselWaves}, the vectorial Bessel beams can be LSE$_{i}$/LSM$_{i}$ with $i=\{x,y\}$ or TE$_z$/TM$_z$, where the subscript denotes the field component that is zero. In the former case, the electric/magnetic transverse field component that is nonzero is set as the scalar Bessel beam solution $U_\nu$ in Eq.~\eqref{eq:ScalarIntegralFormPsi}, while in the latter case, it is the magnetic/electric longitudinal component of the field that is set as $U_\nu$, and the other components are found from Eq.~\eqref{eq:Psi_Gen} via Maxwell equations (see~\hyperref[app:Construction_EM_Bessel_waves]{Supp. Mat. F}).

The LSE$_y$ ($E_y=0, E_x=U_\nu$) and TM$_z$ ($H_z=0, E_z=U_\nu$) solutions are respectively given by (see~\hyperref[app:Construction_EM_Bessel_waves]{Supp. Mat. F})
\begin{subequations}
	\begin{equation}
		\begin{aligned}
			\mathbf{E}(\rho,\phi,z)&=\int_0^{2\pi}d\phi_\text{G}\ \psi_\nu(\phi_\text{G})\\
			&\ \times\Big(\hat{\mathbf{x}}+\tan(\delta)\cos(\phi_\text{G})\hat{\mathbf{z}}\Big),
		\end{aligned}
	\end{equation}
	\begin{equation}
		\begin{aligned}
			\eta \mathbf{H}(\rho,\phi,z)&=\int_0^{2\pi}d\phi_\text{G}\ \psi_\nu(\phi_\text{G})\\
			&\ \times\Big(\hspace{-1mm}-\sin(\delta)\tan(\delta)\sin(\phi_\text{G})\cos(\phi_\text{G})\hat{\mathbf{x}}\\
			&\quad +(\sin(\delta)\tan(\delta)\cos^2(\phi_\text{G})+\cos(\delta))\hat{\mathbf{y}}\\
			&\quad+\sin(\delta)\sin(\phi_\text{G})\hat{\mathbf{z}}\Big),
		\end{aligned}
	\end{equation}
	\label{eq:LSEy_Mode}
\end{subequations}
and
\begin{subequations}
	\begin{equation}
		\begin{aligned}
			\mathbf{E}(\rho,\phi,z)&=\int_0^{2\pi}d\phi_\text{G} \ \psi_\nu(\phi_\text{G})\\
			&\ \times\Big(\cot(\delta)\cos(\phi_\text{G})\hat{\mathbf{x}}+\cot(\delta)\sin(\phi_\text{G})\hat{\mathbf{y}}+\hat{\mathbf{z}}\Big),
		\end{aligned}
	\end{equation}
	\begin{equation}
		\begin{aligned}
			\eta \mathbf{H}(\rho,\phi,z)&=\int_0^{2\pi}d\phi_\text{G}\ \psi_\nu(\phi_\text{G})\\
			&\ \times\Big(\hspace{-1mm}-\csc(\delta)\sin(\phi_\text{G})\hat{\mathbf{x}}+\csc(\delta)\cos(\phi_\text{G})\hat{\mathbf{y}}\Big),
		\end{aligned}
	\end{equation}
	\label{eq:TMz_Mode}
\end{subequations}
where $\eta$ is the impedance of free-space. The most striking difference between the LSE/LSM and TE/TM beams resides in the simplest feature of their respective constituent waves. The former have a linear transverse polarization, while the latter have of a constant transverse magnitude.

A detailed investigation of these solutions reveals that the axicon angle ($\delta$) distinctly affects the LSE/LSM and TE/TM modes (see~\hyperref[app:Compression_EM_Bessel_waves]{Supp. Mat. G}). In both cases, increasing $\delta$ compresses the Bessel ring pattern towards the axis of the beam; however, this variation also breaks the symmetry of the transverse LSE/LSM patterns, even for $\nu=0$, whereas it leaves the TE/TM pattern azimuthally symmetric. It is also interesting to note that, for a small axicon angle, i.e., $\delta\ll{\pi/2}$, the LSE$_y$ modes essentially reduce to their $E_x$ and $H_y$ components, similarly to the scalar form.

\begin{figure}[h]
	\centering
	\includegraphics[scale=1]{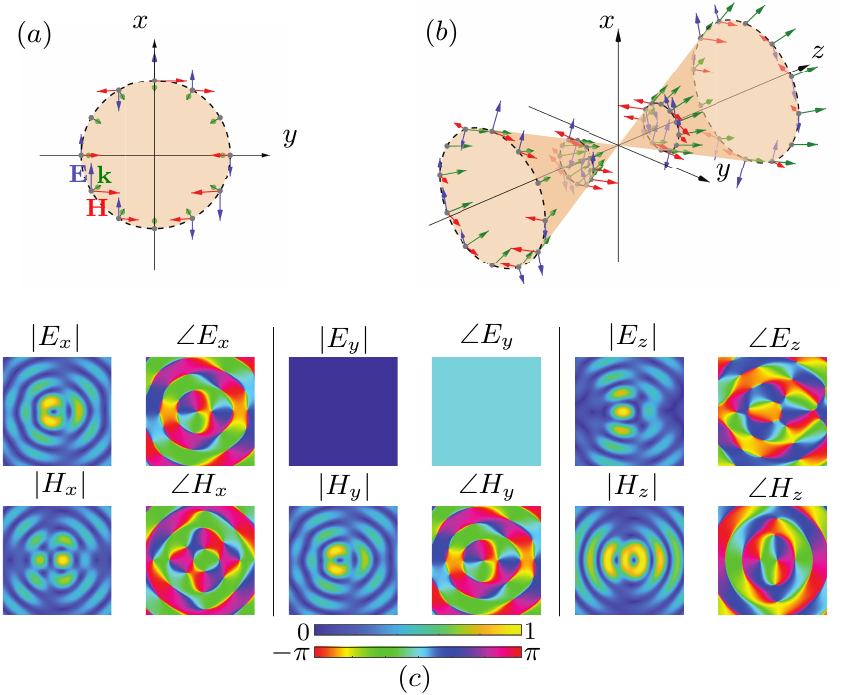}
	\caption{Noninteger global order ($\nu=1.5$) LSE$_{y}$ Bessel beam, computed by Eq.~\eqref{eq:LSEy_Mode}. (a)~Transverse vectorial fields at $z=z_G$ for 12 plane wave samples. (b)~Corresponding complete vectorial fields in 4 different cut planes (dotted circles) with axicon angle $\delta=25^\circ$.  (c)~Magnitude and phase of the fields over the cross-sectional area of $8\lambda\times8\lambda$.}
	\label{fig:LSEy_Cone}
\end{figure}

\begin{figure}[h]
	\centering
	\includegraphics[scale=1]{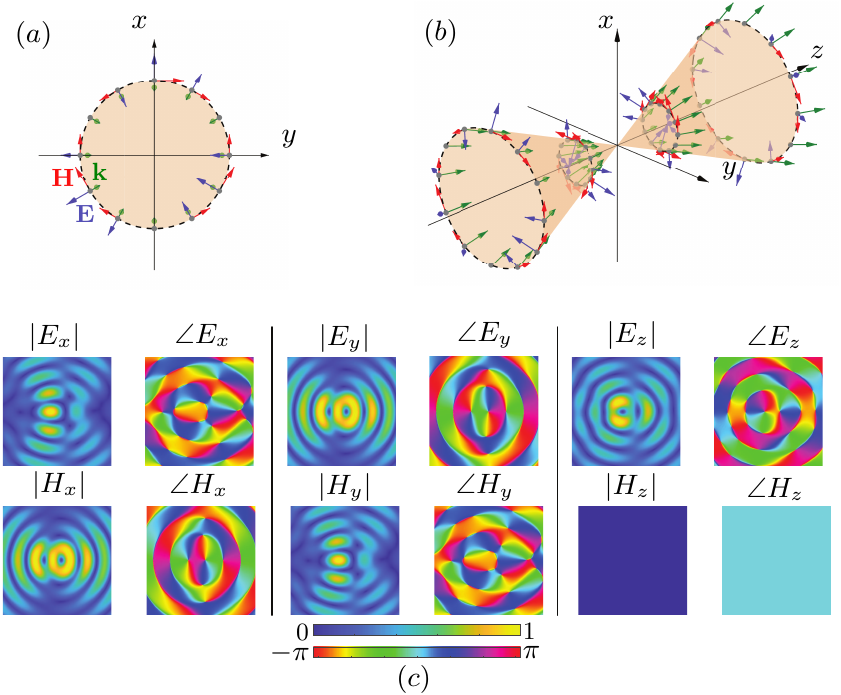}
	\caption{Noninteger global order ($\nu=1.5$) TM$_{z}$ Bessel beam, computed by Eq.~\eqref{eq:TMz_Mode}, with the same parameters and panels as in Fig.~\ref{fig:LSEy_Cone}.}
	\label{fig:TMz_Cone}
\end{figure}

Figures~\ref{fig:LSEy_Cone} and~\ref{fig:TMz_Cone} depict the LSE$_y$ and TM$_z$ Bessel beams of global order $\nu=1.5$ corresponding to the solutions of Eqs.~\eqref{eq:LSEy_Mode} and \eqref{eq:TMz_Mode}, respectively. The vectorial field distributions plotted in the panels (a) and (b) of the two figures represent samples of the constituting waves of the integral construction of the beam (Fig.~\ref{fig:SpectrumDoubleConePart1}). Their strong vectorial nature starkly contrasts with the configuration of the scalar solution, except for the LSE$_y$ case in the aforementioned axicon limit ($\delta\ll{\pi/2}$). Note that the electric field of the constituent waves of the LSE$_y$ mode is linearly polarized in the $x$-direction, while that of the TM$_z$ modes is radially polarized. Nonzero and noninteger $\nu$ vectorial modes are obtained from their fundamental counterpart by simply setting  the $\nu$ parameter in the initial phase of the constituent waves, i.e., $\gamma_\nu(\phi_{\text{G}})$ in Eq.~\eqref{eq:Lin_Phase}, to the desired OAM.

\begin{figure}[h]
	\centering
	\includegraphics[scale=1]{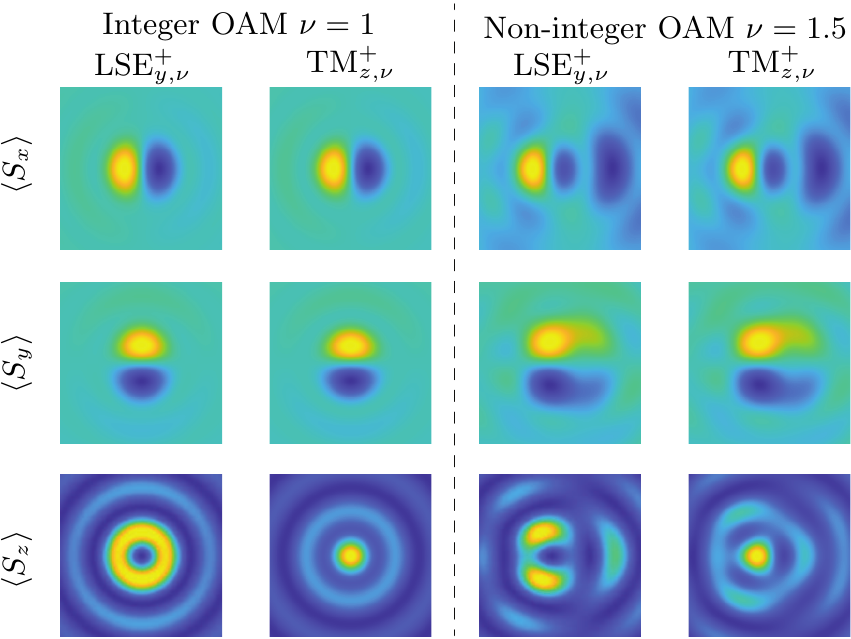}
	\caption{Time-average Poynting vector components of LSE$_y^{+}$ (Eq.~\eqref{eq:LSEy_Mode}) and TM$_z^{+}$ (Eq.~\eqref{eq:LSEy_Mode}) modes with $\delta=25^\circ$.}
	\label{fig:PoyntingComponents}
\end{figure}

Figure~\ref{fig:PoyntingComponents} plots time-average Poynting vectors of integer and non-integer LSE$_y$ and TM$_y$ Bessel beams. Interestingly, whereas the maxima of the LSE$_y$ transverse Poynting vector components are superimposed with those of the longitudinal Poynting vector component, the TM$_z$ transverse maxima are not overlapping the longitudinal maxima. Also notice that the TM$_z$ longitudinal Poynting vector component for $\nu=n=1$ does \emph{not} exhibit a null on the beam axis, contrarily to the case of all nonzero-OAM scalar solutions; this is allowed by the fact that the polarization singularities associated with the radial configuration of the constituent plane waves of the TM/TE modes cancel out the phase singularities in this particular case of $\nu=1$. These various results, with the complementariness of the LSE/LSM--TE/TM modes, and their extension to higher OAMs, illustrate the structural diversity of the vectorial Bessel beams, including horizontal/vertical/right-circular/left-circular LSE/LSM polarization (see Visualizations 1 and 2) and azimuthal/radial/hybrid TE/TM polarization (see Visualization 3), and suggest that they may lead to a wealth of still unexplored opportunities for the optical-force manipulation of nanoparticles.

The LSE/LSM and TM/TE electromagnetic vectorial Bessel beams are related by the following  relations, which may be easily verified upon comparing Eqs.~\eqref{eq:LSEy_Mode} and \eqref{eq:TMz_Mode}:
\begin{subequations}
	\begin{equation}
		\begin{aligned}
			&\frac{1}{2}\left(\begin{Bmatrix}
				\text{LSE}_{y,\nu-1}+\text{LSE}_{y,\nu+1}\\
				\text{LSM}_{y,\nu-1}+\text{LSM}_{y,\nu+1}
			\end{Bmatrix}\right)\\
			&+i\frac{1}{2}\left(\begin{Bmatrix}
				\text{LSE}_{x,\nu-1}-\text{LSE}_{x,\nu+1}\\
				\text{LSM}_{x,\nu-1}-\text{LSM}_{x,\nu+1}
			\end{Bmatrix}\right)=\tan(\delta)\begin{Bmatrix}
				\text{TM}_{z,\nu}\\
				\text{TE}_{z,\nu}
			\end{Bmatrix},
		\end{aligned}
		\label{eq:LSELSMtoTMTE}
	\end{equation}
	
	\begin{equation}
		\begin{aligned}
			&\frac{\tan(\delta)}{2}\left(\begin{Bmatrix}
				\text{TM}_{z,\nu-1}+\text{TM}_{z,\nu+1}\\
				\text{TE}_{z,\nu-1}+\text{TE}_{z,\nu+1}
			\end{Bmatrix}\right)\\
			&\pm i\frac{\sin(\delta)}{2}\left(\begin{Bmatrix}
				\text{TE}_{z,\nu-1}-\text{TE}_{z,\nu+1}\\
				\text{TM}_{z,\nu-1}-\text{TM}_{z,\nu+1}
			\end{Bmatrix}\right)=\begin{Bmatrix}
				\text{LSE}_{y,\nu}\\
				\text{LSM}_{y,\nu}
			\end{Bmatrix},
		\end{aligned}
	\end{equation}
	
	\begin{equation}
		\begin{aligned}
			&\mp \frac{\sin(\delta)}{2}\left(\begin{Bmatrix}
				\text{TE}_{z,\nu-1}+\text{TE}_{z,\nu+1}\\
				\text{TM}_{z,\nu-1}+\text{TM}_{z,\nu+1}
			\end{Bmatrix}\right)\\
			&+ i\frac{\tan(\delta)}{2}\left(\begin{Bmatrix}
				\text{TM}_{z,\nu-1}-\text{TM}_{z,\nu+1}\\
				\text{TE}_{z,\nu-1}-\text{TE}_{z,\nu+1}
			\end{Bmatrix}\right)=\begin{Bmatrix}
				\text{LSE}_{x,\nu}\\
				\text{LSM}_{x,\nu}
			\end{Bmatrix}.
		\end{aligned}
	\end{equation}
\end{subequations}
where the second subscripts ($\nu$ and $\nu\pm{1}$) correspond, as usual, to the global OAM $\nu$ in Eq.~\eqref{eq:Psi_Gen_wave}. Note the OAM conservation between the LSE/LSM and TM/TE modes in each of these relations, and the interesting mediation of the axicon angle. Note also that, in the paraxial approximation ($\delta\ll \pi/2$), the TM/TE beams, with their complex cylindrical (radial/azimuthal/hybrid) polarizations, be realized superpositing two transversally linearly-polarizaed LSE/LSM beams, according to Eq.~\eqref{eq:LSELSMtoTMTE}.

\section{Physical Implementation}\label{sec:phys_impl}

Several techniques have been proposed for generating Bessel beams experimentally. The main ones are axicon lenses illuminated by a Laguerre-Gauss beam~\cite{arlt2000generation}, spatial light modulators~\cite{chattrapiban2003generation}, open circular waveguides with selectively excited modes~\cite{salem2011microwave}, antenna arrays with proper phase feeding network~\cite{lemaitre2012generation}, metasurfaces illuminated by plane waves~\cite{pfeiffer2014controlling}, and 2D circular leaky-wave antennas~\cite{fuscaldo2015higher}. Unfortunately, these techniques are restricted to simple beams, excessively complex to implement, bulky and expensive, or suffering from poor efficiency.

The unified integral formulation presented in this paper (Fig.~\ref{fig:SpectrumDoubleConePart1} with Eq.~\eqref{eq:ScalarIntegralFormPsi} for the scalar case, and Eqs.\eqref{eq:LSEy_Mode} and~\eqref{eq:TMz_Mode} for the vectorial case) naturally points to a generation technique that is immune of these issues and that offers in addition a universal implementation framework. Indeed, circularly distributing a set of sources with the phases, amplitudes and polarizations of the derived modal field solutions (e.g., top panels of Figs.~\eqref{fig:LSEy_Cone} and~\eqref{fig:TMz_Cone}) would exactly and efficiently produce the corresponding Bessel beams, irrespective to their order or complexity.

Specifically, the integral-formulation generation technique consists in the following design steps: 1)~select a sufficient number of sources ($N$) to properly sample the desired OAM according to the Nyquist criterion, 2)~determine an appropriate beam apodization ($w(\xi)$) for each of the constitutent waves to be radiated by these sources, and 3)~adequately set the phase, magnitude and polarization of each of the sources, and orient them so as to launch the constituent waves along a cone with the selected axicon angle ($\delta$). This is mathematically expressed by the formula 
\begin{subequations}
	\begin{equation}
		\mathbf{E}=\sum_{\phi_\text{G}\in[0,2\pi]}^{N} w(\phi_\text{G})\mathbf{E}^{\text{PW}}(\phi_\text{G})\Delta \phi_\text{G},
	\end{equation}
	where $w(\phi_\text{G})$ is the apodization of the constituent waves, $\mathbf{E}^{\text{PW}}(\phi_\text{G})$ is their plane-wave modal field solution (e.g., Eqs.~\eqref{eq:LSEy_Mode} or \eqref{eq:TMz_Mode}), and $\Delta \phi_\text{G}=2\pi/N$. In the case of a (typical) Gaussian apodization, we have 
	\begin{equation}
		w(\phi_\text{G})=e^{-({x_\circ^2(\phi_\text{G})}+{y_\circ^2(\phi_\text{G})})/w_0^2},
	\end{equation}
	where $w_0$ is the waist of the beam, and $(x_\circ(\phi_\text{G}),y_\circ(\phi_\text{G}))$ represents the local conical coordinates
	\begin{align}
		&x_\circ(\phi_\text{G})=\big(x\cos(\phi_\text{G})+y\sin(\phi_\text{G})\big)\cos(\delta)-z\sin(\delta),\\
		&y_\circ(\phi_\text{G})=-x\sin(\phi_\text{G})+y\cos(\phi_\text{G}),
	\end{align}
	\label{eq:Truncated&DiscreteBessel}
\end{subequations}
which are related to the radial conical coordinate $\sqrt{x_\circ(\phi_\text{G})^2+y_\circ(\phi_\text{G})^2}=\xi(\phi_\text{G})$ in $w(\xi)$ (Eq.~\eqref{eq:Psi_Gen_wave}). 

Note that apodization of the plane wave $\mathbf{E}^{\text{PW}}$ by the function $w(\xi)$  results into a localization of the beam in a restricted of extent  $L=w_0/\sin(\delta)$ about the center of the cone at ($z=0$). Moreover, the discretization of the integral induces a distortion of the Bessel pattern, which grows with the distance from the axis of the beam, as previously explained, so that $N$ may have to be increased to provide a satisfactory beam approximation across the transverse area of interest.

\begin{figure}[h]
	\centering
	\includegraphics[scale=1]{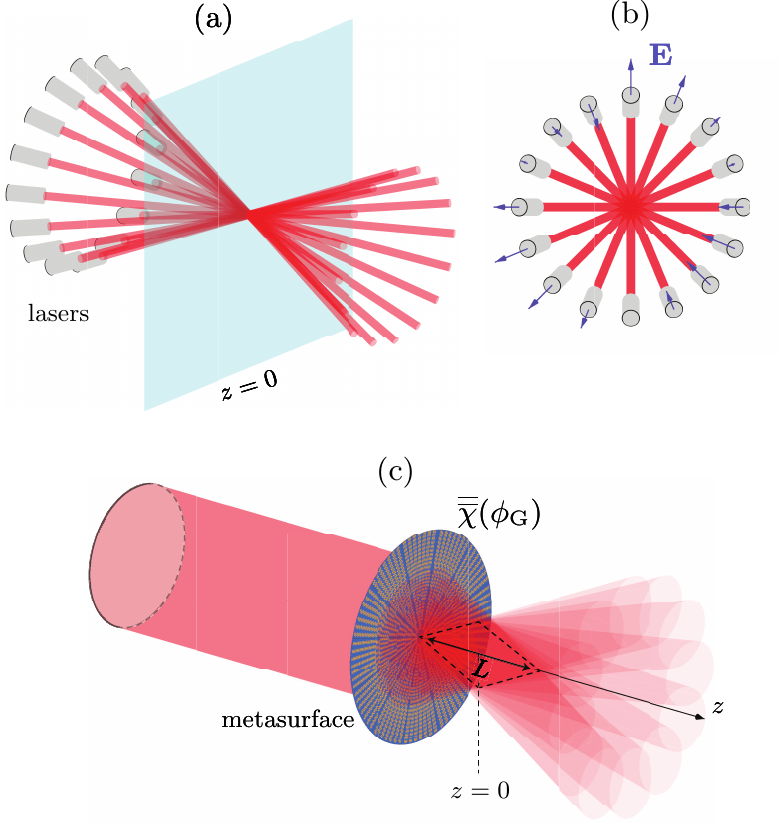}
	\caption{Experimental implementation of the proposed general Bessel beam generation technique. (a)~Using a circular array of $N$ laser sources. (b)~Cross-sectional view of (a) with polarization and initial phase configuration corresponding to the TM$_{z,1.5}$. (c)~Metasurface implementation, with metasurface susceptibility tensor $\overline{\overline{\chi}}(\phi_\text{G})$, with single laser source.}
	\label{fig:Implementations}
\end{figure}

Figure~\ref{fig:Implementations} depicts the experimental implementation of the integral-formulation Bessel beam generation. Figure~\ref{fig:Implementations}(a) represents a direct incarnation of this formulation, which consists of a circular array of laser beams with proper magnitudes, phases and polarizations, as illustrated in Fig.~\ref{fig:Implementations}(b). Such an implementation, involving $N$ independent lasers with respective magnitude, phase and polarization controls, is quite complex and cumbersome. Fortunately, recent advances in metasurface technology suggests the much more practical implementation shown in Fig.~\ref{fig:Implementations}(c). Indeed, this metasurface-based Bessel beam generator requires only one laser source, while being ideally compact and inexpensive.

The metasurface required in the implementation of the Bessel beam generator depicted in Fig.~\ref{fig:Implementations}(c) can be easily realized using latest metasurface synthesis techniques~\cite{achouri2018design}. The simplest implementation strategy would consist 
in cascading metasurfaces that separately tailor the amplitude, the phase, the inclination, and the polarization of the incident wave in the transverse plane of the system. Specifically, assuming a linearly polarized incident wave, such a design would then consist in three cascaded metasurfaces. Two of these metasurfaces would be common to the LSE/LSM and TE/TM cases, with one metasurface providing the required azimuthal phase distribution via azimuthal sectors made of particles inducing progressive transmission delays, and the other providing the required conical inclination via a constant radial phase gradient. In contrast, third metasurface would be different for the LSE/LSM and TE/TM cases. In the former case, given the linear transverse polarization (Eq.~\eqref{eq:LSEy_Mode} and Fig.~\ref{fig:LSEy_Cone}), there is no required polarization processing, and the third metasurface needs to provide the proper transverse magnitude distribution, which can be accomplished with dissipative particles, while in the latter case, given the constant transverse magnitude (Eq.~\eqref{eq:TMz_Mode} and Fig.~\ref{fig:TMz_Cone}), there is no required magnitude processing, and the third metasurface needs to provide the proper transverse polarization distribution, which can be accomplished with birefringent particles.

\section{Conclusion}
We have presented a unified perspective of Bessel beams of arbitrary OAM (zero, integer and noninteger) and nature (scalar, LSE/LSM and TE/TM) based on an integral formulation, and deduced from this formulation a universal and efficient generation technique. The proposed formulation may be extended to other conical beams, such as Mathieu~\cite{gutierrezalternative} and Weber~\cite{bandresparabolic} beams, upon simply adjusting the amplitude modulation function $w(\xi)$ Eq.~\eqref{eq:Psi_Gen_wave} as
\footnote{
		In the case of Mathieu beams, $w(\xi)$ must be replaced by $w_\text{e}(\phi_\text{G})=\text{ce}_m(\phi_\text{G},q)$ and $w_\text{o}(\phi_\text{G})=\text{se}_m(\phi_\text{G},q)$, where $\text{ce}_m(\cdot)$ and $\text{se}_m(\cdot)$ are the even (subscript `e') and odd (subscript 'o') angular Mathieu functions of order $m$ and ellipticity $q$~\cite{guttierrezvegamathieu}, and we set $\gamma_\nu(\phi_\text{G})=0$ in Eq.~\eqref{eq:Lin_Phase}.}
\footnote{In the case of Weber beams, $w(\xi)$ must be replaced by $w_\text{e}(\phi_{\text{G}})=e^{ia\text{ln}\left|\tan(\phi_\text{G}/2)\right|}/\left(2\sqrt{\pi\left| \sin(\phi_\text{G})\right|}\right)$  or $w_\text{o}(\phi_\text{G})=-iw_\text{e}(\phi_\text{G})$ for $0\leq\phi_\text{G}\leq \pi$ and $w_\text{o}(\phi_\text{G})=iw_\text{e}(\phi_\text{G})$ for $\pi<\phi_\text{G}\leq 2\pi$, with $a$ being a parameter, and we set $\gamma_\nu(\phi_\text{G})=0$ in Eq.~\eqref{eq:Lin_Phase}.}, and to nonconical beams, such as the Gauss-Laguerre and Hypergeometric Gaussian beams, upon nesting a corresponding extra integral for the proper spectrum in Eq.~\eqref{eq:ScalarIntegralFormPsi}.
This formulation increases the insight into their characteristics and facilitates their generation for non-complex spectra as conical ones. This global perspective opens up new horizons in structured light for a variety of applications, such as spectroscopy, microscopy, and optical/quantum force manipulations.



{\small
\bibliographystyle{IEEEtran}
\bibliography{PRL_Bessel_Bibliography}}


\clearpage
\addcontentsline{toc}{section}{\bfseries Supplementary Materials}
\renewcommand{\thesubsection}{Supplementary Material \Alph{subsection}}
\setcounter{section}{0}

\setcounter{footnote}{0}  

\renewcommand\thefigure{S\arabic{figure}}
\setcounter{figure}{0}  

\numberwithin{equation}{section}
\renewcommand\theequation{A\arabic{equation}}

\renewcommand{\thesection}{Supplementary Material \Alph{section}}
\section{OAM per Photon for a Bessel Beam of General Order}\label{app:OAM_Bessel_Global_Order}
The \emph{total time-averaged angular momentum (AM)} per photon for a monochromatic vectorial beam with propagation axis $\hat{\mathbf{z}}$ is~\cite{berry1998paraxial}
\begin{equation}
	J_z=\hbar\omega\frac{\int_V\left(\mathbf{r}\times\langle \mathbf{g} \rangle\right)\cdot\hat{\mathbf{z}}dV}{\int_V\langle u \rangle dV},
	\label{eq:OAM_photon_Vect}
\end{equation}
where $\langle \mathbf{g} \rangle=\text{Re}\left\{\mathbf{D}^*\times\mathbf{B}\right\}/2$ is the time-averaged linear electromagnetic momentum density, $\langle u \rangle=\text{Re}\left\{\mathbf{E}^*\cdot\mathbf{D}+\mathbf{B}^*\cdot\mathbf{H}\right\}/4$ is the time-averaged energy density, and $V$ is a cylindrical volume enclosing the beam.

Wave beams, including Bessel beams, are typically satisfactorily described in terms of their \emph{paraxial approximation}. In this regime, the fields $\mathbf{D},\mathbf{B},\mathbf{H}$ can be expressed, from the Maxwell equations and from the constitutive equations, in terms of the field $\mathbf{E}$, and the longitudinal part of $\mathbf{E}$, $E_z$, can be expressed in terms of its transverse part, $\mathbf{E}_\text{T}$. Expressing then $\langle\mathbf{g}\rangle$ and $\langle{u}\rangle$ in Eq.~\eqref{eq:OAM_photon_Vect} in terms of $\mathbf{E}_\text{T}$, neglecting the terms in $1/k^2$ and $1/k^4$~\cite{berry1998paraxial}, and integrating the remaining terms by part yields
\begin{subequations}
	\begin{equation}
		J_z=L_z+S_z,
	\end{equation}
	which splits into
	\begin{align}
		&L_z=\frac{\int_V\text{Re}\left\{-i\hbar\mathbf{E}_\text{T}^*\cdot\big((\mathbf{r}\times\nabla)\cdot\hat{\mathbf{z}}\big)\mathbf{E}_\text{T}\right\}}{\int_V\text{Re}\left\{ \mathbf{E}_\text{T}^*\cdot\mathbf{E}_\text{T}\right\}dV}\label{eq:OAM_per_photon_Paraxial}\\
		\intertext{and}
		&S_z=\frac{\int_V\text{Re}\left\{-i\hbar(\mathbf{E}_\text{T}^*\times\mathbf{E}_\text{T})\cdot\hat{\mathbf{z}}\right\}dV}{\int_V\text{Re}\left\{ \mathbf{E}_\text{T}^*\cdot\mathbf{E}_\text{T}\right\}dV}	
	\end{align}
\end{subequations}
where $L_z$ is the orbital AM (OAM) and $S_z$ is the spin AM (SAM).

In the case of a \emph{Bessel beam with transverse linear polarization} (case of LSE and LSM modes, but not TE and TM modes, in the paper), i.e., $\mathbf{E}_\text{T}=U\hat{\mathbf{x}}$, Eq.~\eqref{eq:OAM_per_photon_Paraxial} reduces to
\begin{equation}
	L_z=\hbar\frac{\text{Im}\left\{\int_0^{2\pi}\int_0^RU^*\frac{\partial U}{\partial \phi}\rho d\rho d\phi\right\}}{\text{Re}\left\{\int_0^{2\pi}\int_0^RU^*U\rho d\rho d\phi\right\}},
	\label{eq:OAM_photon}
\end{equation}
where the radial integrals have been truncated to a finite radius, $R$, to account for the finite energy of a practical beam. 

\begin{figure}[H]
	\centering
	\includegraphics[scale=0.9]{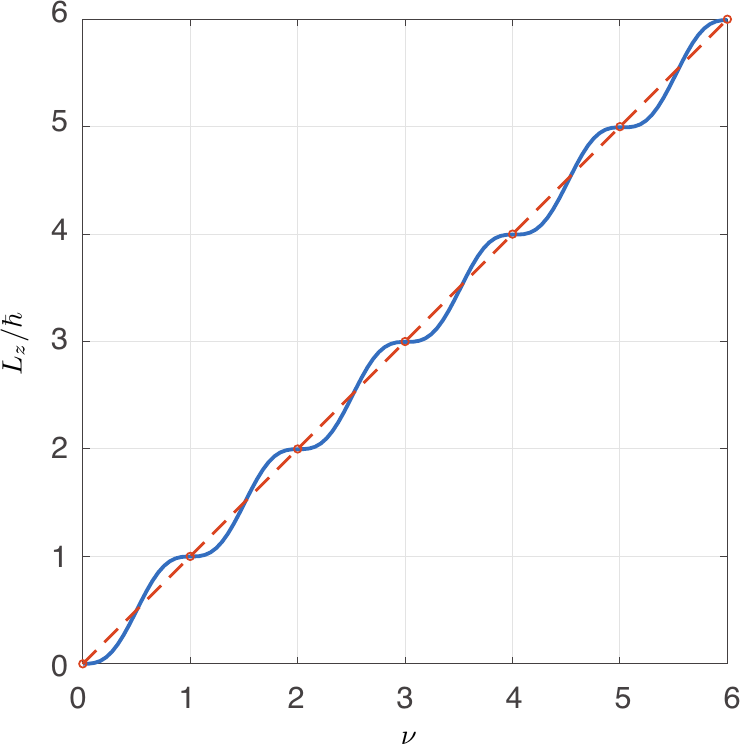}
	\caption{OAM per photon carried by a Bessel beam of order $\nu$ (solid curve) computed by Eq.~\eqref{eq:OAM_photon} with $k=2\pi$, $\delta$ = 25$^\circ$, $\phi_{\text{G},0}=0$ and $R=25$~m, and corresponding linear relation (dashed curve), showing the (nontrivial) relation between the parameter $\nu$ and the OAM ($L_z/\hbar$). The OAM for $\nu\in\mathbb{Z}$ is called the \emph{topological charge}.}
	\label{fig:OAMvsBessel_nu}
\end{figure}

Figure~\ref{fig:OAMvsBessel_nu} plots, using Eq.~\eqref{eq:OAM_photon}, the OAM per photon for a Bessel beam [$U=U_\nu$ in Eq.~(\ref{eq:ScalarIntegralFormPsi}) with $\psi_\nu(\phi_\text{G})$ in Eq.~(\ref{eq:Psi_Gen})] for $R=25$~m. As can be seen, the OAM is a continuous and wiggly function of the parameter $\nu$, which coincides with the linear function $L=\hbar\nu$ only at integer and half-integer values of $\nu$.

\numberwithin{equation}{subsection}
\renewcommand\theequation{B\arabic{equation}}
\section{Verification of the Conical Spatial Angular Spectrum}\label{app:Ang_Spectrum}
We shall show here that the conical spatial angular spectrum
\begin{equation}
	\hspace{-0.5cm}\begin{aligned}
		&\tilde{U}_\nu(k_\rho,k_\phi,k_z)=\\
		&\begin{cases}
			\frac{\delta(k_\rho-\alpha)}{k_\rho}\text{e}^{i\nu\big(k_\phi+(2\pi-(\phi_{\text{G},0}\pm\pi))\big)}\delta(k_z-\beta)\;&\text{if }k_\phi<\phi_{\text{G},0}\pm\pi,\\
			\frac{\delta(k_\rho-\alpha)}{k_\rho}\text{e}^{i\nu\big(k_\phi-(\phi_{\text{G},0}\pm\pi)\big)}\delta(k_z-\beta)\;&\text{if }k_\phi\geq \phi_{\text{G},0}\pm\pi,
		\end{cases}
		\label{SupA:Ang_Spectrum}
	\end{aligned}
\end{equation}
where the $\pm$ solutions in each of the two expressions correspond to the ranges $0\leq \phi_{\text{G},0}\leq\pi$ ($+$ sign) and $\pi\leq \phi_{\text{G},0}<2\pi$ ($-$ sign), which appears in the the spatial inverse Fourier transform
\begin{subequations}
	\begin{equation}
		\begin{aligned}
			U_{\nu}(\rho,\phi,z)&=\mathcal{F}\{\tilde{U}_\nu(k_\rho,k_\phi,k_z)\}\\
			&=\int_0^{\infty}\int_{0}^{2\pi}\int_0^{\infty}\tilde{U}_\nu(k_\rho,k_\phi,k_z)\\
			&\qquad\times \text{e}^{i(\boldsymbol{k}\cdot\mathbf{r}-\omega t)}k_\rho dk_\rho d k_\phi dk_z,\\
		\end{aligned}
	\end{equation}
	where
	\begin{equation}
		\boldsymbol{k}(k_\phi)= k_\rho\cos(k_\phi)\hat{\mathbf{x}}+k_\rho\sin(k_\phi)\hat{\mathbf{y}}+k_z\hat{\mathbf{z}}
		\label{SupB:wavenumber_reciprocal}\footnote{The reader may have noticed that Eq.~(\refeq{SupB:wavenumber_reciprocal}) is different from Eq.~(\ref{eq:Conical_k}), although the two equations fundamentally represent the same entity. Specifically, with $k_z=\beta$ and $k_\phi=\phi_\text{G}$, the transverse components $\mathbf{k}$ have opposite signs in the two equations. This is due to the fact that the convention in Fig.~\ref{fig:SpectrumDoubleConePart1} and Eq.~(\refeq{eq:Conical_k}) is opposite that in the definition  the Fourier transform. Indeed, in the former case, $\mathbf{k}$ points to the spectral origin, whereas in the latter case it points away from it.}
	\end{equation}
	and
	\begin{equation}
		\mathbf{r}(\rho,\phi,z)= \rho\cos(\phi)\hat{\mathbf{x}}+\rho\sin(\phi)\hat{\mathbf{y}}+z\hat{\mathbf{z}},
	\end{equation}
	\label{SupA:Inv_Ftransform}
\end{subequations}
corresponds to the integral representation of Eq.~(\refeq{eq:ScalarIntegralFormPsi}) with Eq.~(\refeq{eq:Psi_Gen}), or Eq.~(\refeq{eq:TransScalInt}).

Substituting Eq.~(\ref{SupA:Ang_Spectrum}) into Eq.~(\ref{SupA:Inv_Ftransform}) and solving the simple integrals in $k_\rho$ and $k_z$ yields

\begin{equation}
	\begin{aligned}
		& U_{\nu}(\rho,\phi,z)=\\
		&\int_{0}^{\phi_{\text{G},0}\pm\pi}\text{e}^{i\big(\alpha\cos(k_\phi-\phi)+\beta z-\omega t\big)}\text{e}^{i\nu\big(k_\phi+(2\pi-(\phi_{\text{G},0}\pm\pi))\big)} d k_\phi\\
		&+\int_{\phi_{\text{G},0}\pm\pi}^{2\pi}\text{e}^{i\big(\alpha\cos(k_\phi-\phi)+\beta z-\omega t\big)}\text{e}^{i\nu\big(k_\phi-(\phi_{\text{G},0}\pm\pi)\big)} d k_\phi.
	\end{aligned}
\end{equation}

The remaining integrals, in $k_\phi$, can then be solved via the change of variables $h_1=k_\phi+(2\pi-(\phi_{\text{G},0}\pm\pi))$ and $h_2=k_\phi-(\phi_{\text{G},0}\pm\pi)$, which leads to
\begin{equation}
	\begin{aligned}
		&U_{\nu}(\rho,\phi,z)=\\
		&\int_{2\pi-(\phi_{\text{G},0}\pm\pi)}^{2\pi}\text{e}^{i\big(-\alpha\cos(h_1+\phi_{\text{G},0}-\phi)+\beta z-\omega{t}\big)}\text{e}^{i\nu h_1} d h_1\\
		&+\int_{0}^{2\pi-(\phi_{\text{G},0}\pm\pi)}\text{e}^{i\big(-\alpha\cos(h_2+\phi_{\text{G},0}-\phi)+\beta z-\omega{t}\big)}\text{e}^{i\nu h_2} d h_2.
		\label{eqAppA:Integralh1h2}
	\end{aligned}
\end{equation}

Then, defining $h_1:=h_0$ and $h_2:=h_0$ transforms this expression into
\begin{equation}
	U_{\nu}(\rho,\phi,z)=\int_{0}^{2\pi}\text{e}^{i\big(-\alpha\cos(h_0+\phi_{\text{G},0}-\phi)+\beta z-\omega t\big)}e^{i\nu\phi_\text{G}} dh_0.
\end{equation}

Finally, using the change of variables $\phi'=h_0+\phi_{\text{G},0}$ transforms this expression into
\begin{equation}
	\begin{aligned}
		&U_{\nu}(\rho,\phi,z)=\\
		&\int_{\phi_{\text{G},0}}^{\phi_{\text{G},0}+2\pi}\text{e}^{i\big(-\alpha\cos(\phi'-\phi)+\beta z-\omega t\big)}e^{i\nu(\phi'-\phi_{\text{G},0})} d\phi',
	\end{aligned}
\end{equation}
which is identical to Eq.~(\ref{eq:TransScalInt}). This proves that $\tilde{U}_\nu(k_\rho,k_\phi,k_z)$ in Eq.~(\ref{SupA:Ang_Spectrum}) is indeed the angular spectrum of $U_{\nu}(\rho,\phi,z)$ in Eq.~(\ref{SupA:Inv_Ftransform}), i.e., that a Bessel beam has an the angular spectrum given by Eq.~(\ref{SupA:Ang_Spectrum}).

\numberwithin{equation}{subsection}
\renewcommand\theequation{C\arabic{equation}}
\section{Derivation of Eq.~(\ref{eq:TransScalInt})}\label{app:Concise_Form}
Substituting Eq.~(\ref{eq:Psi_Gen}) with $w(\xi)=A^{\text{PW}}$ (const.) into Eq.~(\ref{eq:ScalarIntegralFormPsi}) yields
\begin{equation}
	\begin{aligned}
		U_{\nu}(x,y,z,t)
		&=\int_{0}^{2\pi}A^\text{PW}\text{e}^{i(\mathbf{k}(\phi_\text{G})\cdot\mathbf{r}+\gamma_\nu(\phi_\text{G})-\omega t)}d\phi_\text{G},
	\end{aligned}
	\label{eq:general_integral}
\end{equation}
where, assuming $0<\phi_\text{G}<2\pi$,
\begin{equation}
	\gamma_\nu(\phi_\text{G})=\begin{cases}
		\nu\big( \phi_\text{G}+(2\pi-\phi_{\text{G},0}) \big) & 0\le\phi_\text{G}<\phi_{\text{G},0},\\
		\nu(\phi_\text{G}-\phi_{\text{G},0}) & \phi_{\text{G},0}\le\phi_\text{G}<2\pi.
	\end{cases}
	\label{eq:gam_nu}
\end{equation}

Inserting the Eq.~(\ref{eq:gam_nu}) into Eq.~(\ref{eq:general_integral}) splits the integral of the former as
\begin{equation}
	\begin{aligned}
		U_{\nu}(x,y,z,t)=A^\text{PW}\int_{0}^{\phi_{\text{G},0}}\text{e}^{i(\mathbf{k}(\phi_\text{G})\cdot\mathbf{r}+\nu\big( \phi_\text{G}+(2\pi-\phi_{\text{G},0}) \big)-\omega t)}d\phi_\text{G}\\
		+A^\text{PW}\int_{\phi_{\text{G},0}}^{2\pi}\text{e}^{i(\mathbf{k}(\phi_\text{G})\cdot\mathbf{r}+\nu(\phi_\text{G}-\phi_{\text{G},0})-\omega t)}d\phi_\text{G}.
	\end{aligned}
\end{equation}

Operating next the change of variable $\phi_1=\phi_\text{G}+2\pi$  in the first integral transforms this relation into
\begin{equation}
	\begin{aligned}
		U_{\nu}(x,y,z,t)=A^\text{PW}\int_{2\pi}^{2\pi+\phi_{\text{G},0}}\text{e}^{i(\mathbf{k}(\phi_1)\cdot\mathbf{r}+\nu(\phi_1-\phi_{\text{G},0})-\omega t)}d\phi_1\\
		+A^\text{PW}\int_{\phi_{\text{G},0}}^{2\pi}\text{e}^{i(\mathbf{k}(\phi_\text{G})\cdot\mathbf{r}+\nu(\phi_\text{G}-\phi_{\text{G},0})-\omega t)}d\phi_\text{G}.
	\end{aligned}
\end{equation}

Finally, the two integrals in this relation can be grouped into a single one via the definitions $\phi_1:=\phi'$ and $\phi_\text{G}:=\phi'$, which results into
\begin{equation}
	U_{\nu}(x,y,z,t)=A^\text{PW}\int_{\phi_{\text{G},0}}^{\phi_{\text{G},0}+2\pi}\text{e}^{i(\mathbf{k}(\phi')\cdot\mathbf{r}+\nu(\phi'-\phi_{\text{G},0})-\omega t)}d\phi'.
\end{equation}

\numberwithin{equation}{subsection}
\renewcommand\theequation{D\arabic{equation}}
\section{Derivation of Eq.~(\ref{eq:Complex_Orthogonal_Expansion})}\label{app:Orthogonal_Expansion}
As any function, the (nonperiodic) complex exponential function $\text{e}^{i\nu\phi'}$ with $\nu\in\mathbb{R}\backslash\mathbb{Z}$ and domain $\left[\phi_{\text{G},0},\phi_{\text{G},0}+2\pi\right[$\footnote{The description of the nonperiodic function $\text{e}^{i \nu\phi'}$ over the domain $\phi'\in\left[\phi_{\text{G},0},\phi_{\text{G},0}+2\pi\right[$ is equivalent to describing this nonperiodic  function as a piecewise function in terms of $\phi_\text{G}\in\left[0,2\pi\right[$ where conditions depend on $\phi_{\text{G},0}$.} in Eq.~(\ref{eq:TransScalInt}) can be decomposed over a complete orthogonal basis. Choosing for this basis the periodic complex exponential function $\text{e}^{im\phi'}$ ($m\in\mathbb{Z}$) leads then to the expansion
\begin{equation}
	\text{e}^{i\nu\phi'}=\sum_{m=-\infty}^{+\infty}c_m\text{e}^{im\phi'}.
	\label{eqAppB:OrthoExpansionGeneral}
\end{equation}
Multiplying both sides of this relation by $\text{e}^{-im'\phi'}$ and integrating over one period of the function $\text{e}^{im\phi'}$ yields
\begin{equation}
	\begin{aligned}
		\int_{\phi_{\text{G},0}}^{\phi_{\text{G},0}+2\pi}\text{e}^{i\nu\phi'}\text{e}^{-im'\phi'}d\phi'
		=\int_{\phi_{\text{G},0}}^{\phi_{\text{G},0}+2\pi}\sum_{m=-\infty}^{+\infty}c_m\text{e}^{-i(m'-m)\phi'}d\phi'.
	\end{aligned}
\end{equation}
Inverting the order of the integral and the sum in the right-handside term and using the orthogonality property of the integer-order comlex integral function gives then
\begin{equation}
	\int_{\phi_{\text{G},0}}^{\phi_{\text{G},0}+2\pi}\text{e}^{i(\nu-m')\phi'}d\phi'=2\pi c_{m'},
\end{equation}
which integrates into
\begin{equation}
	c_m= \frac{\sin\big((m-\nu)\pi\big)}{(m-\nu)\pi}\text{e}^{-i(m-\nu)({\phi_{\text{G},0}}+\pi)},
	\label{eqAppB:OrthoExpansionCoefficient}
\end{equation}
where the dummy prime has been dropped.

Substituting Eq.~(\ref{eqAppB:OrthoExpansionCoefficient}) into Eq.~(\ref{eqAppB:OrthoExpansionGeneral}) finally yields the explicit expansion
\begin{equation}
	\text{e}^{i\nu \phi'}=\sum_{m=-\infty}^{+\infty}\text{sinc}\big((m-\nu)\pi\big)\text{e}^{-i(m-\nu)(\phi_{\text{G},0}+\pi)}\text{e}^{i m \phi'}.
	\label{eqAppB:Complex_Orthogonal_Expansion}
\end{equation}

\numberwithin{equation}{subsection}
\renewcommand\theequation{E\arabic{equation}}
\section{Physical Interpretation of the Phase Parameter $\phi_{\text{G},0}$}
\label{app:NL_Discontinuity_Parameter}

The physical role of the phase term $\phi_{\text{G},0}$ in Eq.~\eqref{eq:Lin_Phase} and  Eq.~\eqref{eq:BesselAnyOAM} is not trivial. This section clarifies this issue, using two distinct and complementary strategies: 1)~analyzing the problem in the spatial-frequency spectral ($\mathbf{k}$) domain, where it is simpler, and 2)~inspecting a beam made of truncated constituent waves in a cross-sectional plane where these waves weakly overlap and therefore only locally (i.e., only with nearest neighbors, see Fig.~\ref{fig:SpectrumDoubleConePart1}) interfere, which might be considered as a spatial alternative to 1).

\begin{figure}[H]
	\centering
	\includegraphics[scale=1]{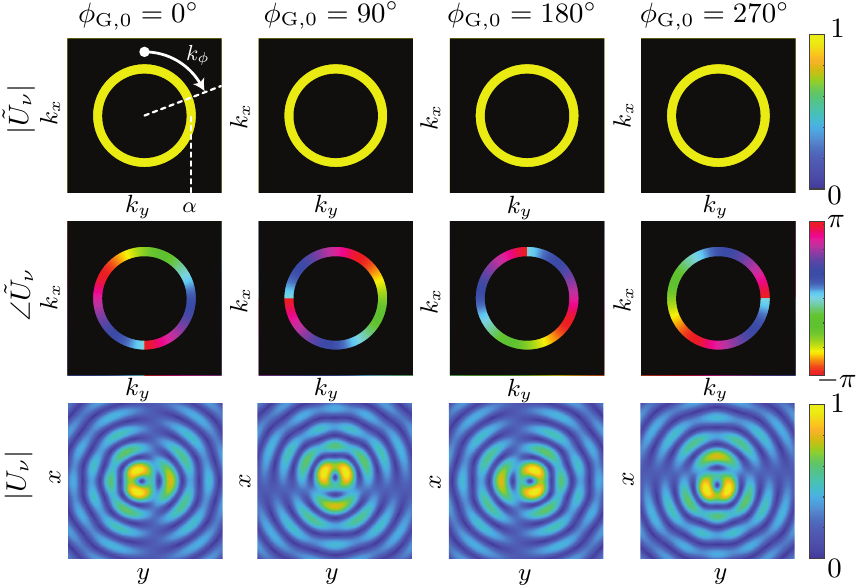}
	\caption{Effect of $\phi_{\text{G},0}$ on the spectrum of a Bessel beam [Eq.~\eqref{SupE:Ang_SpectrumPhiG0}] formed by plane constituent waves with $\nu=1.5$ and $\delta$ = 25$^\circ$. In the first two rows, we have introduced a non-zero width to circular section of the cone for visualization. The bottom row shows, for comparison, the amplitude of the direct field, obtained by inverse-Fourier transforming Eq.~\eqref{SupE:Ang_SpectrumPhiG0} and represented at the apex of the cone ($z=0$).}
	\label{fig:SpectrumPhaseSingularityPosition}
\end{figure}

The spatial-frequency spectrum solution of the general solution in Eq.~(\refeq{eq:BesselAnyOAM}) is (see {Supp. Mat. B})
\begin{equation}
	\hspace{-0.5cm}\begin{aligned}
		&\tilde{U}_\nu(k_\rho,k_\phi,k_z)=\\
		&\begin{cases}
			\frac{\delta(k_\rho-\alpha)}{k_\rho}\text{e}^{i\nu\big(k_\phi+(2\pi-(\phi_{\text{G},0}\pm\pi))\big)}\delta(k_z-\beta)\;&\text{if }k_\phi<\phi_{\text{G},0}\pm\pi,\\
			\frac{\delta(k_\rho-\alpha)}{k_\rho}\text{e}^{i\nu\big(k_\phi-(\phi_{\text{G},0}\pm\pi)\big)}\delta(k_z-\beta)\;&\text{if }k_\phi\geq \phi_{\text{G},0}\pm\pi.
		\end{cases}
		\label{SupE:Ang_SpectrumPhiG0}
	\end{aligned}
\end{equation}
Equation~(\refeq{SupE:Ang_SpectrumPhiG0}) describes a cone with different initial phases in the $\mathbf{k}$-space. If $\nu=n\in\mathbb{Z}$, the term $\text{e}^{in2\pi}=1$ in the top expression disappears so that this expression reduces to the second; as a result, the initial phase $\phi_{\text{G},0}\pm\pi$ does not depend any more on $k_\phi$, and represents therefore an dummy initial phase that can be set to zero, which reduces the spectrum to $\tilde{U}_n(k_\rho,k_\phi,k_z)=(\delta(k_\rho-\alpha)/k_\rho)\text{e}^{in{k_\phi}}\delta(k_z-\beta)$. If $\nu\in\mathbb{R}\backslash\mathbb{Z}$, the situation is obviously more complex, with a phase ($k_\phi$) discontinuity of $2\pi\nu$ appearing in Eq.~\eqref{SupE:Ang_SpectrumPhiG0}, specifically
\begin{equation}
	\lim_{\varepsilon\rightarrow 0}\angle\tilde{U}_\nu(k_\phi=\phi_{\text{G},0}\pm\pi)-\angle\tilde{U}_\nu(k_\phi=\phi_{\text{G}}\pm\pi-\varepsilon)=2\pi\nu.
\end{equation}
Thus, the parameter $\phi_{\text{G},0}$ corresponds then the position of the spectral phase discontinuity as $k_{\phi,0}=\phi_{\text{G},0}+\pi$. Figure~\ref{fig:SpectrumPhaseSingularityPosition} illustrates these results, 

\begin{figure}[H]
	\centering
	\includegraphics[scale=1]{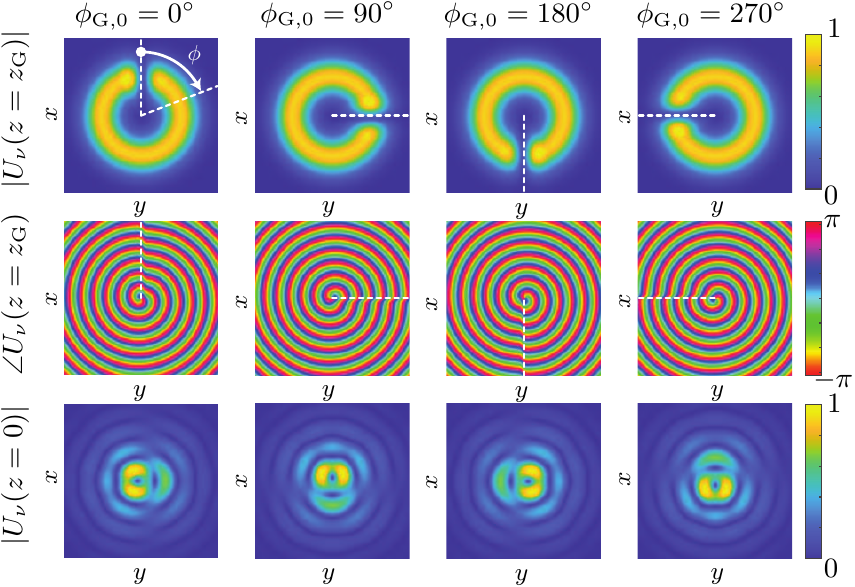}
	\caption{Complex amplitude and phase of the direct fields associated with a Bessel beam with Gaussian circular-cylindrically truncated constituting in a plane of overlap only the nearest neighbours ($z_\text{G}<z<0$) and in the overlaping plane ($z=0$) (same parameters as in Fig.~\ref{fig:SpectrumPhaseSingularityPosition}).}
	\label{fig:TruncatedPhaseSingularityPosition}
\end{figure}

Moving now on to the second strategy, Fig.~\ref{fig:TruncatedPhaseSingularityPosition} plots the effect of $\phi_{\text{G},0}$ on the direct fields associated with Bessel beam with circular-cylindrically truncated constituting waves. Here, the parameter $\phi_{\text{G},0}$, has also an impact of the phase, with the neighboring waves canceling each other at the angle $\phi=\phi_{\text{G},0}$ ($z<0$). 

\begin{figure}[H]
	\centering
	\includegraphics[scale=1]{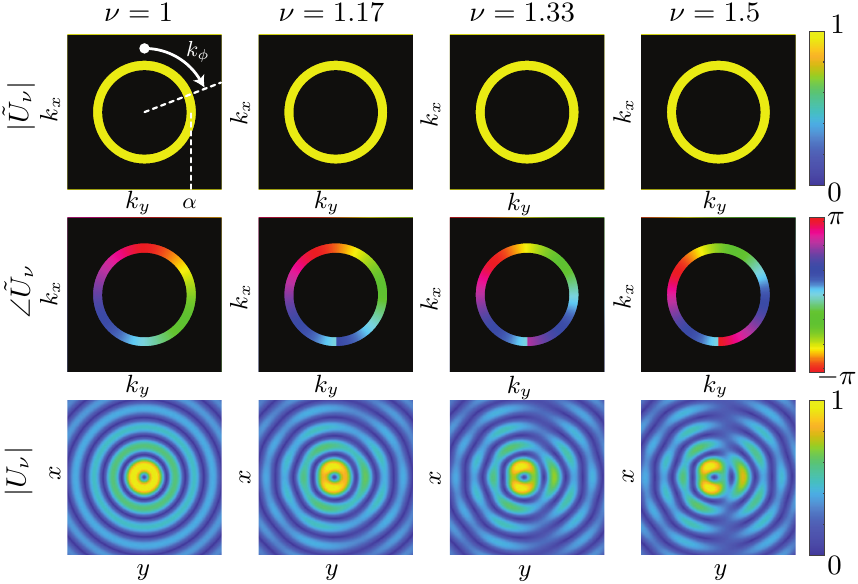}
	\caption{Same parameters as in Fig.~\ref{fig:SpectrumPhaseSingularityPosition} with different global order $\nu$.}
	\label{fig:SpectrumPhaseSingularityJump}
\end{figure}

\begin{figure}[H]
	\centering
	\includegraphics[scale=1]{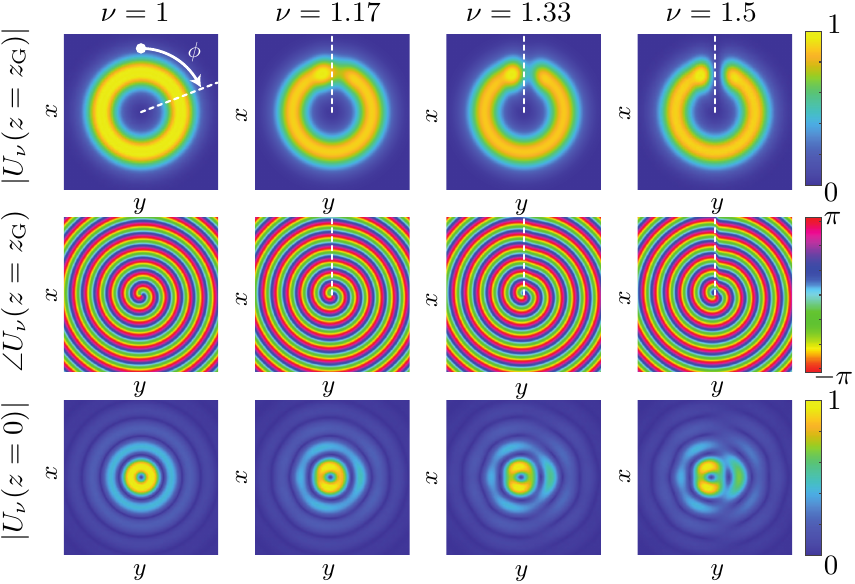}
	\caption{Same parameters as in Fig.~\ref{fig:TruncatedPhaseSingularityPosition} with different global order $\nu$.}
	\label{fig:TruncatedPhaseSingularityJump}
\end{figure}

Figures~\ref{fig:SpectrumPhaseSingularityJump}  and~\ref{fig:TruncatedPhaseSingularityJump} show the effect of the phase discontinuity, as the Figures~\ref{fig:SpectrumPhaseSingularityPosition}  and~\ref{fig:TruncatedPhaseSingularityPosition}  respectively, for a given $\phi_{\text{G,0}}=0$ and different global order $\nu$. Figure~\ref{fig:TruncatedPhaseSingularityJump} shows how the amplitude (of neighboring waves) opening is affected by the phase discontinuity $2\pi\nu$.

\numberwithin{equation}{subsection}
\renewcommand\theequation{F\arabic{equation}}
\section{Construction of the Electromagnetic Vectorial Bessel Waves}\label{app:Construction_EM_Bessel_waves}

\subsection{Maxwell Equations for a Plane Wave}
The free-space Maxwell equations for a harmonic plane wave propagating in the direction $\mathbf{k}$ read
\begin{subequations}
	\begin{align}
		\begin{split}
			\eta \mathbf{H}^{\text{PW}}=\frac{1}{k}\big((&k_yE_z^{\text{PW}}-k_zE_y^{\text{PW}})\hat{\mathbf{x}}+(-k_xE_z^{\text{PW}}+k_zE_x^{\text{PW}})\hat{\mathbf{y}}\\
			&+(k_xE_y^{\text{PW}}-k_yE_x^{\text{PW}})\hat{\mathbf{z}}\big),\label{eq:MaxwellEq1}
		\end{split}\\
		\begin{split}
			\mathbf{E}^{\text{PW}}=\frac{\eta}{k}\big((-&k_yH_z^{\text{PW}}+k_zH_y^{\text{PW}})\hat{\mathbf{x}}+(k_xH_z^{\text{PW}}-k_zH_x^{\text{PW}})\hat{\mathbf{y}}\\
			&+(-k_xH_y^{\text{PW}}+k_yH_x^{\text{PW}})\hat{\mathbf{z}}\big),\label{eq:MaxwellEq2}
		\end{split}\\
		k_x&E_x^{\text{PW}}+k_yE_y^{\text{PW}}+k_zE_z^{\text{PW}}=0,\label{eq:MaxwellEq3}\\
		k_x&H_x^{\text{PW}}+k_yH_y^{\text{PW}}+k_zH_z^{\text{PW}}=0,\label{eq:MaxwellEq4}
	\end{align}
\end{subequations}
where $\eta=\sqrt{\mu_0/\epsilon_0}$ is the impedance of free-space and  $k=\omega\sqrt{\mu_0\epsilon_0}$ is the wavenumber of free-space, with $\epsilon_0=8.854\cdot{10}^{-12}$~F/m and $\mu_0=1.257\cdot{10}^{-6}$~N/A$^2$.

\subsection{Construction of the LSE/LSM Modes}
Let us start with the LSE$_y$ ($E_y=0$) modes. In this case, we have
\begin{equation}
	E_y^{\text{PW}}(\phi_\text{G})=0,
	\label{eq:EyvsphiG0}
\end{equation}
from which we find, upon solving Eq.~(\ref{eq:MaxwellEq3}) for $E_z^{\text{PW}}$,
\begin{subequations}
	\begin{equation}
		E_z^{\text{PW}}(\phi_\text{G})=-\frac{k_x(\phi_\text{G})}{k_z(\phi_\text{G})}E_x^{\text{PW}}(\phi_\text{G}),
	\end{equation}
	which becomes, upon setting $E_x^{\text{PW}}(\phi_\text{G})=\psi_\nu(\phi_\text{G})$ (transverse field in the scalar case) in Eq.~\eqref{eq:Psi_Gen_wave} and using Eq.~(\ref{eq:Conical_k}),
	\begin{equation}
		E_z^{\text{PW}}(\phi_\text{G})=\tan(\delta)\cos(\phi_\text{G})\psi_\nu(\phi_\text{G}).
		\label{eq:EzvsphiG}
	\end{equation}
\end{subequations}
The corresponding components of $\mathbf{H}^{\text{PW}}$ are found by inserting Eqs.~(\ref{eq:EyvsphiG0}) and~(\ref{eq:EzvsphiG}) into Eq.~(\ref{eq:MaxwellEq1}), and successively solving for the different components, which results into
\begin{subequations}
	\begin{align}
		\eta H_x^{\text{PW}}(\phi_\text{G})&=\left(-\sin(\delta)\tan(\delta)\sin(\phi_\text{G})\cos(\phi_\text{G})\right)\psi_\nu(\phi_\text{G}),\\
		\eta H_y^{\text{PW}}(\phi_\text{G})&=\left( \sin(\delta)\tan(\delta)\cos^2(\phi_\text{G})+\cos(\delta)\right)\psi_\nu(\phi_\text{G}),\\
		\eta H_z^{\text{PW}}(\phi_\text{G})&=\sin(\delta)\sin(\phi_\text{G})\psi_\nu(\phi_\text{G}).
	\end{align}
\end{subequations}

The LSE$_y$ Bessel beam corresponding to the integral form of Eq.~(\ref{eq:ScalarIntegralFormPsi}) is then formed, upon grouping the previous results, by the field components
\begin{subequations}
	\begin{align}
		E_x&=\int_0^{2\pi}d\phi_\text{G} \ \psi_\nu(\phi_\text{G}),\\
		E_y&=0,\\
		E_z&=\int_0^{2\pi}d\phi_\text{G} \ \psi_\nu(\phi_\text{G})\tan(\delta)\cos(\phi_\text{G}), \\
		\eta H_x&=\int_0^{2\pi}d\phi_\text{G} \ \psi_\nu(\phi_\text{G})\big(-\sin(\delta)\tan(\delta)\sin(\phi_\text{G})\cos(\phi_\text{G})\big),\\
		\eta H_y&=\int_0^{2\pi}d\phi_\text{G} \ \psi_\nu(\phi_\text{G})\big( \sin(\delta)\tan(\delta)\cos^2(\phi_\text{G})+\cos(\delta) \big),\\
		\eta H_z&=\int_0^{2\pi}d\phi_\text{G} \ \psi_\nu(\phi_\text{G})\sin(\delta)\sin(\phi_\text{G}).
	\end{align}
\end{subequations}

The LSE$_x$ Bessel beam field components are similarly found as
\begin{subequations}
	\begin{align}
		E_x&=0\\
		E_y&=\int_0^{2\pi}d\phi_\text{G} \ \psi_\nu(\phi_\text{G})\\
		E_z&=\int_0^{2\pi}d\phi_\text{G} \ \psi_\nu(\phi_\text{G})\tan(\delta)\sin(\phi_\text{G})
	\end{align}
	\begin{align}
		\eta H_x&=\int_0^{2\pi}d\phi_\text{G} \ \psi_\nu(\phi_\text{G})\big( -\sin(\delta)\tan(\delta)\sin^2(\phi_\text{G})-\cos(\delta) \big)\\
		\eta H_y&=\int_0^{2\pi}d\phi_\text{G} \ \psi_\nu(\phi_\text{G})\sin(\delta)\tan(\delta)\sin(\phi_\text{G})\cos(\phi_\text{G})\\
		\eta H_z&=\int_0^{2\pi}d\phi_\text{G} \ \psi_\nu(\phi_\text{G})\big(-\sin(\delta)\cos(\phi_\text{G})\big)
	\end{align}
\end{subequations}
For the sake of completeness, we finally give the LSM$_x$ and LSM$_y$ modes, which are respectively
\begin{subequations}
	\begin{align}
		E_x&=\int_0^{2\pi}d\phi_\text{G} \ \psi_\nu(\phi_\text{G}) \big( \sin(\delta)\tan(\delta)\sin^2(\phi_\text{G})+\cos(\delta) \big),\\
		E_y&=\int_0^{2\pi}d\phi_\text{G} \ \psi_\nu(\phi_\text{G})\big(-\sin(\delta)\tan(\delta)\sin(\phi_\text{G})\cos(\phi_\text{G})\big),\\
		E_z&=\int_0^{2\pi}d\phi_\text{G} \ \psi_\nu(\phi_\text{G})\sin(\delta)\cos(\phi_\text{G}),
	\end{align}
	\begin{align}
		\eta H_x&=0,\\
		\eta H_y&=\int_0^{2\pi}d\phi_\text{G} \ \psi_\nu(\phi_\text{G}),\\
		\eta H_z&=\int_0^{2\pi}d\phi_\text{G} \ \psi_\nu(\phi_\text{G})\tan(\delta)\sin(\phi_\text{G}).
	\end{align}
\end{subequations}
and
\begin{subequations}
	\begin{align}
		E_x&=\int_0^{2\pi}d\phi_\text{G} \ \psi_\nu(\phi_\text{G})\sin(\delta)\tan(\delta)\sin(\phi_\text{G})\cos(\phi_\text{G})\\
		E_y&=\int_0^{2\pi}d\phi_\text{G} \ \psi_\nu(\phi_\text{G})\big(-\sin(\delta)\tan(\delta)\cos^2(\phi_\text{G})-\cos(\delta) \big)\\
		E_z&=\int_0^{2\pi}d\phi_\text{G} \ \psi_\nu(\phi_\text{G})\big(-\sin(\delta)\sin(\phi_\text{G})\big)
	\end{align}
	
	\begin{align}
		\eta H_x&=\int_0^{2\pi}d\phi_\text{G} \ \psi_\nu(\phi_\text{G})\\
		\eta H_y&=0\\
		\eta H_z&=\int_0^{2\pi}d\phi_\text{G} \ \psi_\nu(\phi_\text{G})\tan(\delta)\cos(\phi_\text{G})
	\end{align}
\end{subequations}
\subsection{Construction of the TM/TE Modes}
Let us start with the TM$_z$ ($H_z=0$) modes. In this case, we have
\begin{equation}
	H_z^{\text{PW}}(\phi_\text{G})=0.
	\label{eq:HzvsphiG}
\end{equation}
Additionally, we find from the $z$-component of Eq.~(\ref{eq:MaxwellEq1}) the following relation between the transverse components of the electric field:
\begin{subequations}
	\begin{equation}
		E_x^{\text{PW}}(\phi_\text{G})=\frac{k_x(\phi_\text{G})}{k_y(\phi_\text{G})}E_y^{\text{PW}}(\phi_\text{G})=\cot(\phi_\text{G})E_y^{\text{PW}}(\phi_\text{G}),
		\label{eq:trans_ExfunctionEy}
	\end{equation}
	or
	\begin{equation}
		E_y^{\text{PW}}(\phi_\text{G})=\frac{k_y(\phi_\text{G})}{k_x(\phi_\text{G})}E_x^{\text{PW}}(\phi_\text{G})=\tan(\phi_\text{G})E_x^{\text{PW}}(\phi_\text{G}),
		\label{eq:trans_EyfunctionEx}
	\end{equation}
	\label{eq:trans_ExyfunctionEyx}
\end{subequations}
where Eq.~(\ref{eq:Conical_k}) has been used in the last equalities. We find then, upon successively solving Eq.~(\ref{eq:MaxwellEq3}) for $E_x^{\text{PW}}$ and $E_y^{\text{PW}}$, and respectively using Eq.~\eqref{eq:trans_EyfunctionEx} and Eq.~\eqref{eq:trans_ExfunctionEy},
\begin{subequations}
	\begin{equation}
		\begin{aligned}
			E_x^{\text{PW}}(\phi_\text{G})&=\frac{-k_x(\phi_\text{G})k_z(\phi_\text{G})}{k_x^2(\phi_\text{G})+k_y^2(\phi_\text{G})}E_z^{\text{PW}}(\phi_\text{G})\\
			&=\cot(\delta)\cos(\phi_\text{G})E_z^{\text{PW}}(\phi_\text{G}),
		\end{aligned}
	\end{equation}
	or
	\begin{equation}
		\begin{aligned}
			E_y^{\text{PW}}(\phi_\text{G})&=\frac{-k_y(\phi_\text{G})k_z(\phi_\text{G})}{k_x^2(\phi_\text{G})+k_y^2(\phi_\text{G})}E_z^{\text{PW}}(\phi_\text{G})\\
			&=\cot(\delta)\sin(\phi_\text{G})E_z^{\text{PW}}(\phi_\text{G}),
		\end{aligned}
	\end{equation}
	which becomes, upon setting $E_z^{\text{PW}}(\phi_\text{G})=\psi_\nu(\phi_\text{G})$ (transverse field in the scalar case) in Eq.~\eqref{eq:Psi_Gen_wave} and using Eq.~(\ref{eq:Conical_k}),
	\begin{equation}
		E_x^{\text{PW}}(\phi_\text{G})=\cot(\delta)\cos(\phi_\text{G})\psi_\nu(\phi_\text{G}),
		\label{eq:ExvsphiG}
	\end{equation}
	or
	\begin{equation}
		E_y^{\text{PW}}(\phi_\text{G})=\cot(\delta)\sin(\phi_\text{G})\psi_\nu(\phi_\text{G}).
		\label{eq:EyvsphiG}
	\end{equation}
\end{subequations}

The corresponding components of $\mathbf{H}^{\text{PW}}$ are found by inserting Eqs.~(\ref{eq:ExvsphiG}) and ~(\ref{eq:EyvsphiG}) into Eq.~(\ref{eq:MaxwellEq1}), and successively solving for the different components, which results into
\begin{align}
	\eta H_x^{\text{PW}}(\phi_\text{G})&=\left(-\csc(\delta)\sin(\phi_\text{G})\right)\psi_\nu(\phi_\text{G}),\\
	\eta H_y^{\text{PW}}(\phi_\text{G})&=\csc(\delta)\cos(\phi_\text{G})\psi_\nu(\phi_\text{G}),\\
	\eta H_z^{\text{PW}}(\phi_\text{G})&=0.
\end{align}
The TM$_z$ Bessel beam corresponding to the integral form of Eq.~(\ref{eq:ScalarIntegralFormPsi}) is then formed, upon grouping the previous results, by the field components
\begin{subequations}
	\begin{align}
		E_x&=\int_0^{2\pi}d\phi_\text{G} \ \psi_\nu(\phi_\text{G})\cot(\delta)\cos(\phi_\text{G}),\\
		E_y&=\int_0^{2\pi}d\phi_\text{G} \ \psi_\nu(\phi_\text{G})\cot(\delta)\sin(\phi_\text{G}),\\
		E_z&=\int_0^{2\pi}d\phi_\text{G} \ \psi_\nu(\phi_\text{G}),
	\end{align}
	
	\begin{align}
		\eta H_x&=\int_0^{2\pi}d\phi_\text{G} \ \psi_\nu(\phi_\text{G})\big(-\csc(\delta)\sin(\phi_\text{G})\big),\\
		\eta H_y&=\int_0^{2\pi}d\phi_\text{G} \ \psi_\nu(\phi_\text{G})\csc(\delta)\cos(\phi_\text{G}),\\
		\eta H_z&=0.
	\end{align}
\end{subequations}
The TE$_z$ Bessel beam integral field components are similarly found as
\begin{subequations}
	\begin{align}
		E_x&=\int_0^{2\pi}d\phi_\text{G} \ \psi_\nu(\phi_\text{G})\csc(\delta)\sin(\phi_\text{G})\\
		E_y&=\int_0^{2\pi}d\phi_\text{G} \ \psi_\nu(\phi_\text{G})\big(-\csc(\delta)\cos(\phi_\text{G})\big)\\
		E_z&=0
	\end{align}
	
	\begin{align}
		\eta H_x&=\int_0^{2\pi}d\phi_\text{G} \ \psi_\nu(\phi_\text{G})\cot(\delta)\cos(\phi_\text{G})\\
		\eta H_y&=\int_0^{2\pi}d\phi_\text{G} \ \psi_\nu(\phi_\text{G})\cot(\delta)\sin(\phi_\text{G})\\
		\eta H_z&=\int_0^{2\pi}d\phi_\text{G} \ \psi_\nu(\phi_\text{G})
	\end{align}
\end{subequations}

\numberwithin{equation}{subsection}
\renewcommand\theequation{F\arabic{equation}}
\section{Effect of the Axicon Angle on the Transverse Pattern}\label{app:Compression_EM_Bessel_waves}

\begin{figure}[H]
	\centering
	\includegraphics[scale=1]{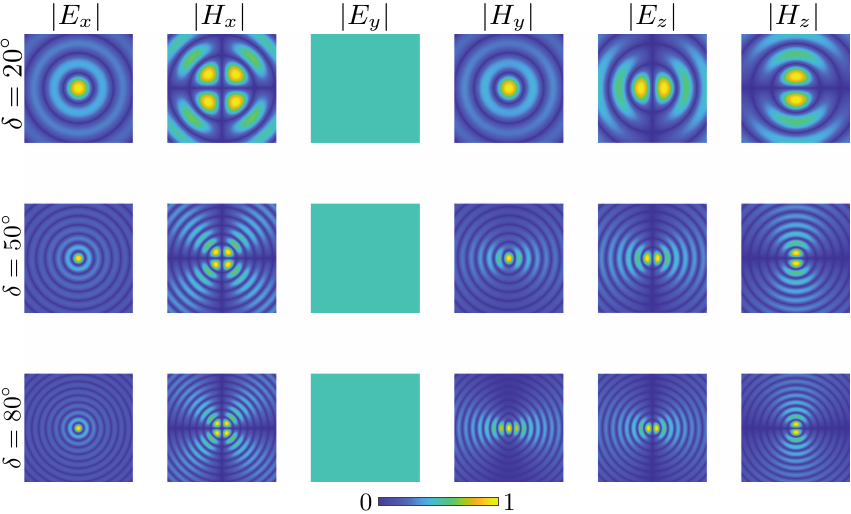}
	\caption{Cross-sectional view of the complex fields amplitude of the LSE$_{y,0}$ Bessel beam, computed by Eq.~\eqref{eq:ScalarIntegralFormPsi} with~\eqref{eq:Psi_Gen_wave} with $500$ constituent plane waves and for different $\delta$ angles, with $\nu=0$, $w(\xi)=1$, $\phi_{\text{G},0}=0$, and over the cross-sectional area of $4\lambda\times 4\lambda$ ($\lambda=2\pi c/\omega$) co-centered with the axis of the beam.}
	\label{fig:LSEyCompression}
\end{figure}

\begin{figure}[H]
	\centering
	\includegraphics[scale=1]{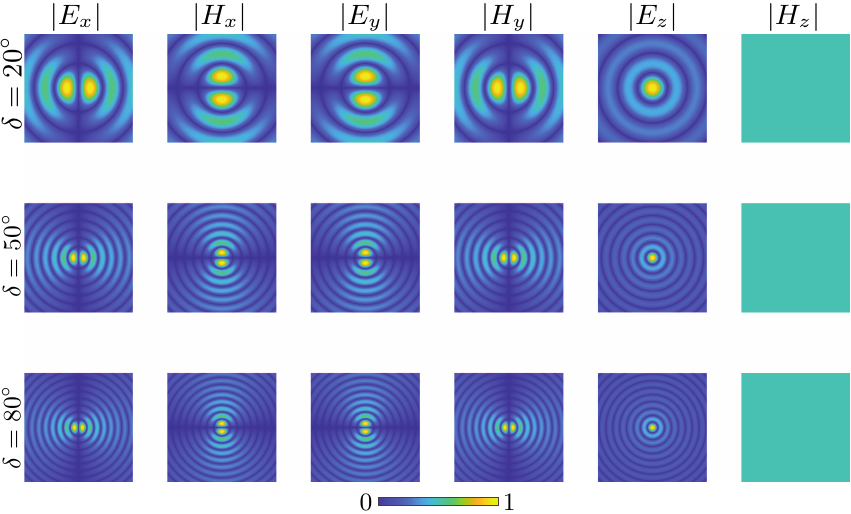}
	\caption{Same as in Fig.~\ref{fig:LSEyCompression} but for the mode TM$_{z,0}$.}
	\label{fig:TMzCompression}
\end{figure}

Figures~\ref{fig:LSEyCompression} and~\ref{fig:TMzCompression} show the effect the axicon angle ($\delta$) on the amplitude of the transverse fields. 

Essentially, increasing $\delta$ increases the transverse spatial frequency $\alpha$ (see Eqs.~\eqref{eq:Conical_k} and~\eqref{eq:BesselAnyOAMa}) of the Bessel waveform of each of the field components and hence compresses the Bessel ring pattern toward the beam axis. 

\begin{figure}[H]
	\centering
	\includegraphics[scale=1]{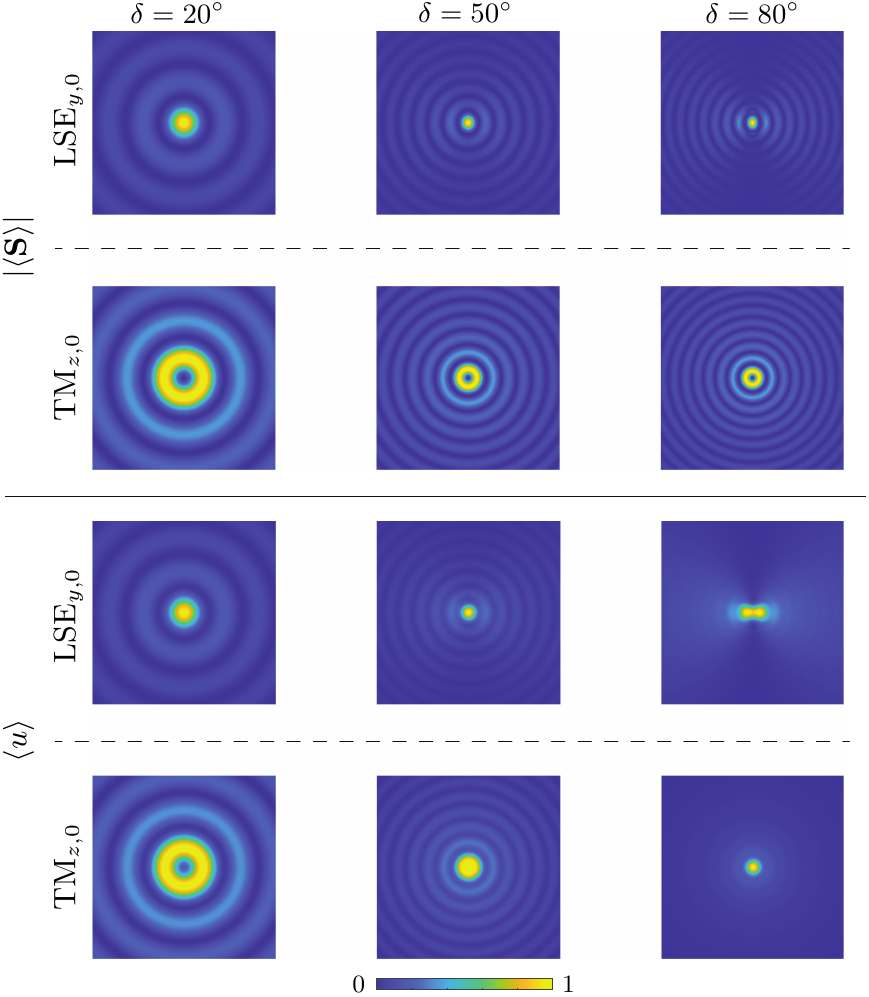}
	\caption{Norm of the Poynting vector and energy density op the LSE$_{y,0}$ and  TM$_{z,0}$ modes for parameters as in Fig.~\ref{fig:LSEyCompression}.}
	\label{fig:EnergyCompression}
\end{figure}

Figure~\ref{fig:EnergyCompression} shows the effect of $\delta$ on the norm of the time-averaged Poynting vector and time-averaged energy density $\langle u\rangle$: Increasing $\delta$ breaks the azimuthal symmetry of the Poynting vector and the energy density of the LSE$_y$ mode, but not that of the TM$_z$ mode, as could be expected from the previous resutls.

\numberwithin{equation}{subsection}
\renewcommand\theequation{G\arabic{equation}}
\section{Comparison between this approach and the approach in~\cite{chafiqoptical}}\label{app:BBs_vs_BBswithGBenv}

The recently published paper~\cite{chafiqoptical}, entitled ``Optical Fourier transform of pseudo-nondiffracting beams'', includes a fundamental similarity but also fundamentals differences with respect to this paper in the manner the beams are mathematically constructed. The two contributions may therefore be considered as complementary to each other. 

\subsection{Similarity}
The fundamental similarity is that both papers describe the localized (or non-diffractive) beam in terms of its angular spectrum decomposition, as opposed to a direct representation of the mathematical function of the beam, which allows a more general and deeper description.

\subsection{Differences}

\subsubsection{Plane vs Gaussian Constituent Waves}

\begin{figure}[ht]
	\centering
	\includegraphics[scale=0.32]{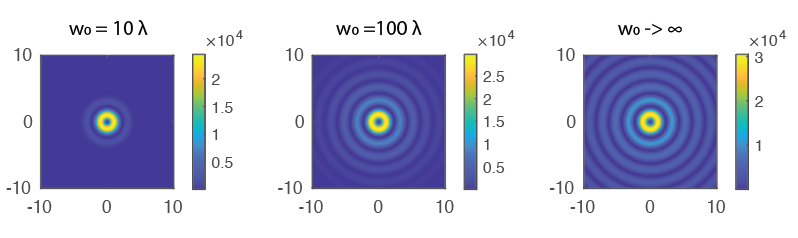}
	\caption{Intensity of the scalar field ($|U_1|^2$) with Gaussian constituent beams of different waist sizes, $w_0$, with $\delta=15^\circ$, computed by Eq.~(10). The limit case $w_0\rightarrow\infty$ naturally corresponds to plane waves.}
	\label{fig:GBtoPW}
\end{figure}

Both papers describe the beam as a superposition of waves over its conical angular spectrum. However, the analytical expressions in this paper are based on a superposition of plane waves, whereas~\cite{chafiqoptical} uses a superposition of Gaussian beams. Since planes waves cannot be implemented in practice, the former approach might a priori appear unpractical. However, the plane wave approach offers invaluable advantages, which will be emphasized in Sec.~B2, without compromising practicality. Indeed, typical Gaussian beams, as produced by lasers, have a waist that is more than three orders of magnitude larger than the operating optical wavelength ($w_0\ggg\lambda$), so that they may be perfectly approximated by plane waves in the interference zone of the beam. This is illustrated by Fig.~\ref{fig:GBtoPW}, which shows that a Bessel beam formed by constituent Gaussian beams of more than 100$\lambda$ waist ($w_0>100\lambda$) cannot be distinguished from one formed by constituent planes waves in the region of interest. Finally, note that Gaussian constituent beams can \emph{numerically} tested in this paper, if desired, although they do not accommodate analytical formulas.

\subsubsection{Vectorial vs Scalar Beams}\label{subsec:vec_scal_cf}

The reason why the formulation of~\cite{chafiqoptical} allows analytical expressions despite the use of constituent Gaussian beams if the fact that is based on the paraxial approximation. In contrast, the formulation of this paper is fully vectorial, and therefore provides a much more advanced description of the beam:

\paragraph{High Axicon Angle}
It is valid for arbitrarily high axicon angles, as required in the case of highly focalized beams, whereas the formulation of~\cite{chafiqoptical} is inapplicable in this regime;

\paragraph{Complex Polarization}
It allows 1)~to distinguish the LSE and LSM modes, 2)~to describe the radially polarized TM$_0$ and azimuthally polarized TE$_0$ modes, and all the higher-order hybridally polarized TE and TM modes, and 3)~to understand complex beam features such as non-integer OAMs, whereas the formulation of~\cite{chafiqoptical} is restricted to an approximate linearily polarized global `mode' with purely integer OAM.

\end{document}